\documentclass[useAMS,usenatbib]{mn2e}
\usepackage[pdftex]{graphicx}
   \usepackage{epstopdf}
\usepackage{graphicx}

\title[HST Optical Transmission Spectroscopy of the Exoplanet HD
189733\lowercase{b}]{Hubble Space Telescope Transmission
  Spectroscopy of the Exoplanet HD 189733\lowercase{b}: 
  High-altitude atmospheric haze in the optical and near-UV with STIS }

\author[D. K. Sing, F. Pont, et al.]
{D. K. Sing$^{1}$\thanks{E-mail: sing@astro.ex.ac.uk}, F. Pont$^{1}$,
S. Aigrain$^{2}$, 
D. Charbonneau$^{3}$, 
J.-M. D\'{e}sert$^{3}$,
N. Gibson$^{2}$,\newauthor
R. Gilliland$^{4}$,  
W. Hayek$^{1}$,
G. Henry$^{5}$,
H. Knutson$^{6}$, 
A. Lecavelier des Etangs$^{7}$,\newauthor
T. Mazeh$^{8}$, A. Shporer$^{9, 10}$\\
$^{1}$Astrophysics Group, School of Physics, University of Exeter, Stocker
  Road, Exeter, EX4 4QL\\ 
$^{2}$Department of Physics, University of Oxford, Denys Wilkinson Building, Keble Road, Oxford OX1 3RH\\
$^{3}$Harvard-Smithsonian Center for Astrophysics, Cambridge, MA 02138\\
$^{4}$Space Telescope Science Institute, 3700 San Martin Drive, Baltimore, MD 21218\\
$^{5}$Tennessee State University, Center of Excellence in Information Systems, 3500 John A. Merritt Blvd., P.O. Box 9501, Nashville, \\TN 37209, USA\\
$^{6}$Department of Astronomy, University of California, Berkeley, 601 Campbell Hall, Berkeley, CA 94720, USA\\
$^{7}$Institut d'Astrophysique de Paris, CNRS; Universit\'{e} Pierre et Marie Curie, 98 bis bv Arago, F- 75014 Paris, France\\
$^{8}$School of Physics and Astronomy, Raymond and Beverly Sackler Faculty of Exact Sciences, Tel Aviv University, Tel Aviv 69978, Israel\\
$^{9}$Las Cumbres Observatory Global Telescope Network, 6740 Cortona Drive, Suite 102, Santa Barbara, CA 93117, USA\\
$^{10}$Department of Physics, Broida Hall, University of California, Santa Barbara, CA 93106, USA\\
}
\begin{document}

\date{Accepted 2011 May 26. Received 2011 May 17; in original form 2011 February 22}

\bibliographystyle{mn2e}

\pagerange{\pageref{firstpage}--\pageref{lastpage}} \pubyear{2011}

\maketitle

\label{firstpage}

\begin{abstract}
We present {\it Hubble Space Telescope} optical and near-ultraviolet
transmission spectra of the transiting hot-Jupiter HD189733b, taken
with the repaired Space Telescope Imaging
Spectrograph (STIS) instrument.  The resulting spectra cover the range 
2900-5700~\AA\ and reach per-exposure signal-to-noise levels greater than 11,000 within a 500~\AA\ bandwidth.
We used time series spectra obtained during two transit events to determine the wavelength dependance of the
planetary radius and measure the exoplanet's atmospheric transmission spectrum for the first time over this wavelength range.  Our
measurements, in conjunction with existing HST spectra, now provide a
broadband transmission spectrum covering the full optical regime.
The STIS data also shows unambiguous evidence of a large occulted stellar spot 
during one of our transit events, which we use to
place constraints on the characteristics of the K dwarf's stellar spots,
estimating spot temperatures around T$_{\mathrm{eff}}\sim$4250~K.
With contemporaneous ground-based photometric monitoring of the stellar
variability, we also 
measure the correlation between the stellar activity level and
transit-measured planet-to-star radius contrast, which is in good agreement with predictions.
We find a planetary transmission spectrum in good agreement with
that of Rayleigh scattering from a high-altitude atmospheric haze as previously
found from HST ACS camera. 
The high-altitude haze is now found to cover the entire optical
regime and is well characterised by Rayleigh scattering.  These
findings suggest that haze may be a globally
dominant atmospheric feature of the planet which would result in a high optical
albedo at shorter optical wavelengths.
\end{abstract}


\section{Introduction}

Transiting planets now allow the possibility of detecting and studying
extrasolar planets, from their formation and statistical properties,
to the bulk composition of the planets themselves as well as their atmospheres.
Atmospheric information on exoplanets can be gathered in several
different ways.
During a transit, an exoplanet passes in front of its host
star, with some of the stellar light filtered through the planet's
atmosphere, making it possible to perform transmission spectroscopy \citep{2002ApJ...568..377C, 
  2003Natur.422..143V}.  
Optical and infrared measurements during secondary eclipse can be used to obtain an exoplanet's
dayside emission spectra, where such properties including the temperature, thermal structure,
and composition can be measured (e.g. \citealt{
2005Natur.434..740D, 
2008Natur.456..767G, 
2008ApJ...686.1341C, 
2009ApJ...707.1707R, 
2009A&A...493L..31S}). 
Finally, orbital phase curves can probe the global temperature distribution and
study atmospheric circulation \citep{2007Natur.447..183K}. 

Transiting events around bright stars are extremely valuable, and have allowed unprecedented access to detailed
atmospheric studies of extrasolar planets.  Two particularly
favorable cases, the hot Jupiters HD~209458b and HD~189733b, have been
the focus of many followup studies and modeling efforts over the last decade.
The first exoplanet atmosphere was detected by
\cite{2002ApJ...568..377C} 
using transits of HD~209458b with the HST STIS instrument.
Sodium was later confirmed from ground-based spectrographs 
which measured similar absorption levels as the revised STIS data
\citep{2008A&A...487..357S, 2008ApJ...686..658S}. 
HD~209458b is now known to have a very low optical albedo
\citep{2006ApJ...646.1241R}. 
The overall optical transmission spectrum is consistent with 
H$_2$ Rayleigh scattering and absorption from Na, and perhaps TiO/VO 
\citep{2008ApJ...686..667S, 2008A&A...485..865L, 2008A&A...492..585D}.
The sodium line profile also reveals the planet's thermosphere and
ionization layer at high altitudes \citep{2011A&A...527A.110V}. 
HD~209458b also shows evaporation of atomic H, C, O, and Si
\citep{2003Natur.422..143V, 2004ApJ...604L..69V, 2010ApJ...717.1291L} 
through ultraviolet observations.
In the infrared, CO has been detected
from high-resolution spectrographs \citep{2010Natur.465.1049S}. 
From secondary eclipse measurements, the planet also has a well
established thermal inversion layer \citep{2008ApJ...673..526K,
2007ApJ...668L.171B}.  

Contrary to HD~209458b, the other well studied hot-Jupiter HD~189733b does not
show spectral features in emission that would indicate a thermal
inversion layer \citep{2008Natur.456..767G,
  2008ApJ...686.1341C}. 
In addition, orbital phase curve measurements have revealed efficient
redistribution via eastward jets, producing small day/night temperature
contrasts \citep{2007Natur.447..183K, 2009ApJ...690..822K}. 
In the optical, a ground-based detection of sodium has been made from transmission spectroscopy by \cite{
2008ApJ...673L..87R},  
with the wider overall optical spectrum consistent with a high altitude haze \citep{
2008MNRAS.385..109P, 
2009A&A...505..891S}. 
The optical transmission spectrum of the haze is seen to become gently
more transparent from blue to red wavelengths, and is consistent with Rayleigh scattering by small condensate
particles, with MgSiO$_3$ grains suggested as a possible candidate \citep{2008A&A...481L..83L}. 
Ultraviolet transits have also determined the planet to have an escaping atomic hydrogen 
atmosphere 
\citep{2010A&A...514A..72L}. 
The infrared opacity is now known to include H$_2$O absorption, with
the best current evidence coming from secondary eclipse measurements
\citep{2008Natur.456..767G}, 
while CO and CO$_2$ have also been suggested
\citep{2009ApJ...699..478D, 2009ApJ...690L.114S}. 

For highly irradiated planets, the atmosphere at optical wavelengths
is vital to the energy budget of the planet, as it is where the bulk 
of the stellar flux is deposited.  The atmospheric properties
at these wavelengths are directly linked to important global properties
such as atmospheric circulation, thermal inversion layers,
and inflated planetary radii.   Some of these features may be linked
within the hot-Jupiter exoplanetary class
(e.g. \citealt{2008ApJ...678.1419F}).  
High-S/N broadband transmission spectra of exoplanets with HST Space
Telescope Imaging Spectrograph (STIS) can provide direct observational constraints
and fundamental measurements of the atmospheric properties
at these vital wavelengths.  This is especially crucial in the near-UV and
into the optical, given that JWST will not be able to observe at such wavelengths.

Here we present Hubble Space Telescope ({\it HST}) transmission spectra of HD~189733b
using the STIS instrument, part of programme GO-11740 aiming to
provide a reference measurement of the broadband transmission spectrum
across the whole visible and near-infrared spectral range.
Our observational strategy in this program is a detailed study for one
exoplanet, in order to gain robust and highly constraining measurements, useful for comparative planetology.
In the case of HD~189733b, this is particularly important as this
planet has become a canonical hot-Jupiter along with HD~209458b, with
both planets providing the foundation for what is now known about these types of
exoplanets.  The STIS instrument, recently repaired during the servicing
mission four (SM4), has once again
provided the opportunity to make extremely high S/N optical transit
observations, capable of detecting and scrutinizing atmospheric constituents.
In this paper, we present new HST/STIS observations, constraining the
atmospheric transmission spectrum in the optical and near-UV.

\section{Observations}
\subsection{Hubble Space Telescope STIS Spectroscopy}
We observed transits of HD~189733b with the HST STIS G430L
grating during 20 November 2009 and 18 May 2010 as part of program GO
11740 (P.I. FP).  Similar observations to these have been made previously on
the other bright hot-Jupiter exoplanet HD~209458b 
\citep{
2007ApJ...655..564K, 
2007Natur.445..511B,
2008ApJ...686..658S}. 
The G430L dataset 
consists of about 200 spectra spanning the two transit events covering
the wavelength range 2,892-5,700~{\AA}, with a resolution R of
$\lambda$/$\Delta\lambda=$530--1,040 ($\sim$2 pixels; 5.5~{\AA}),  
and taken with a wide 52''$\times$2'' slit to minimize slit light
losses.  
This observing technique has proven to
produce high signal-to-noise (S/N) spectra which are photometrically accurate near the 
Poisson limit during a transit event.
Both visits of HST were scheduled such that the third orbit
of the spacecraft would contain the transit event, providing good
coverage between second and third contact as well as an out-of-transit
baseline time series before and after the transit. 
Exposure times of 64 seconds were used in conjunction with a 128-pixel wide
sub-array, which reduces the readout time between exposures
to 23 seconds, providing a 74\% overall duty cycle.

\begin{figure}
\includegraphics[width=0.475\textwidth]{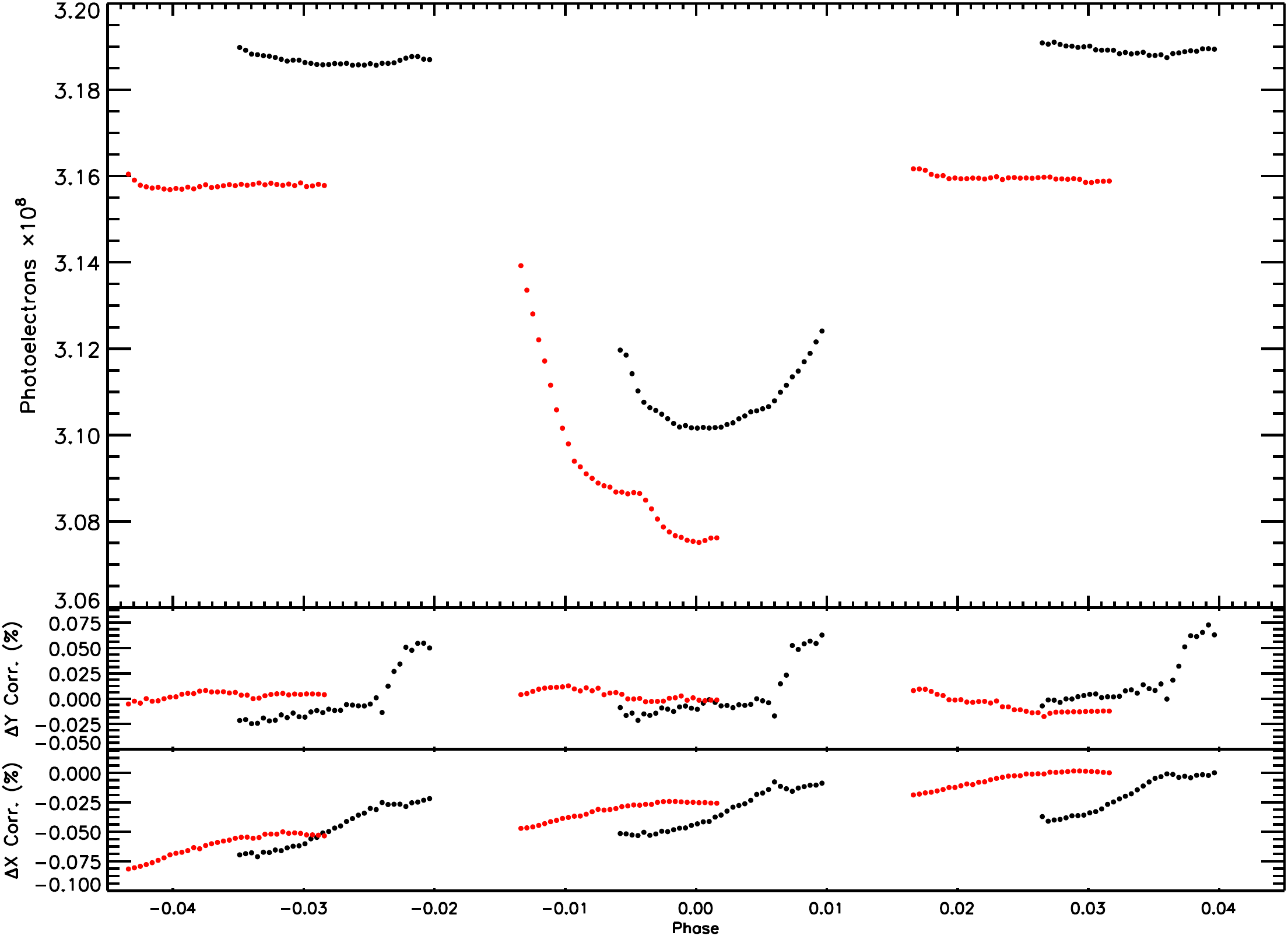}
\caption{(Top) Raw STIS white light curve for visit 1 (black) and
  visit 2 (red).  Three HST spacecraft orbits per visit are plotted with the
  transit visible in the middle orbit.  The bottom two plots show the detector position
systematic trends used to correct the photometric timeseries in
conjunction with the HST orbital phase.}
\label{figwhite}
\end{figure}

The dataset was pipeline-reduced with the latest version of CALSTIS 
and cleaned for cosmic ray detections before performing spectral 
extractions. The aperture extraction was done with IRAF using a
13-pixel-wide aperture, background subtraction, and 
no weights applied to the aperture sum. 
The extracted spectra were then Doppler-corrected to a common rest frame through cross-correlation, 
which removed sub-pixel shifts in the dispersion direction. 
The STIS spectra were then used to create both a white-light photometric 
time series (see Fig. \ref{figwhite}), as well as five wavelength bands covering the G430L spectra, 
integrating the appropriate wavelength flux  from each exposure for
the different bandpasses.  The resulting photometric light curves
exhibit all the expected systematic instrumental effects taken during similar
high S/N transit observations before HST servicing mission four, as originally noted
in \cite{2001ApJ...553.1006B}. 
The main instrument-related systematic effect is primarily due to the well known
thermal breathing of HST, which warms and cools the telescope during the 96 minute 
day/night orbital cycle, causing the focus to vary\footnote{see STScI
  Instrument Science Report ACS 2008-03}.  Previous observations have
shown that once the telescope is slewed to a new pointing position,
it takes approximately one spacecraft orbit to thermally relax, which
compromises the photometric stability of the first orbit of each HST visit.  In
addition, the first exposure of each spacecraft orbit is found to be 
significantly fainter than the remaining exposures.  These trends are continued
in our post-SM4 STIS observations, and are both minimized in the analysis with proper HST visit scheduling.  Similar to other studies, in our
subsequent analysis of the transit we discarded the first orbit
of each visit (purposely scheduled well before the transit event) and
discarded the first exposure of every orbit.
During the first orbit we performed a linearity test on the CCD,
alternating between sets of 60 and 64 second exposures, which confirmed the
linearity of the detector to the high S/N levels used in this program.

As part of a separate HST program aimed at studying atmospheric sodium (GO 11572, P.I. DKS) we also
obtained similar high S/N transits of HD~189733b with the G750M
grating which covers the spectral range between 5813-6382~\AA.  The details of these
observations will be given in a forthcoming paper (Huitson, Sing, et al. in prep) but some of the observations and results are used here for sections \ref{SecERAC}.

\subsection{Stellar Activity Monitoring}

The star HD 189733 is an active K dwarf with a rotation period of about 12 days and a flux variable at the 1-2 percent level. Monitoring the behaviour 
of the activity around the time of the HST visit is essential to obtain the maximum amount of information from the STIS spectroscopic time series.

\begin{figure}
\includegraphics[width=0.475\textwidth]{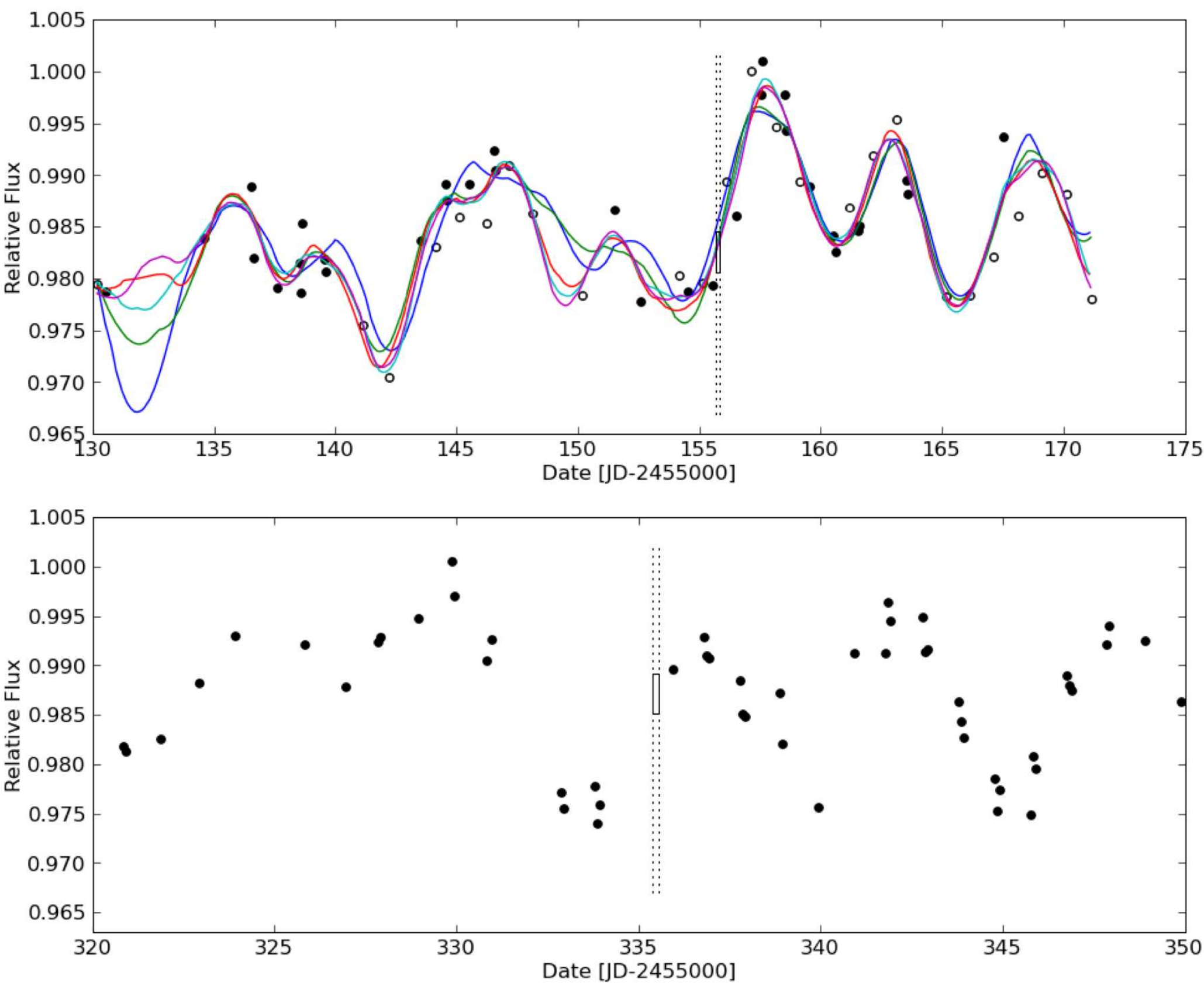}
\caption{APT (filled circles) and Wise (open circles) photometry for
  HD~189733 for HST visit 1 (top) and visit 2 (bottom).  Visit 1 has five representative
  many-spot variability models overplotted.  The HST dates of both visits are indicated with the
  dashed lines, along with the estimated stellar flux levels at that epoch (boxes).}
\label{figwise}
\end{figure}

The ground-based coverage was provided by the T10 0.8 m Automated
Photoelectric Telescope (APT) at Fairborn Observatory in southern
Arizona and spans all of our HST visits (see Fig.~\ref{figwise}).  This ongoing observing campaign of HD189733 began in October
2005 and is detailed in \cite{2008AJ....135...68H}.  
The APT uses two
photomultiplier tubes to simultaneously gather Stromgren $b$ and $y$
photometry.  
This dataset also covers the HD189733 stellar activity during the
epochs of the HST transmission spectra from
\cite{2007A&A...476.1347P}, \cite{2008Natur.452..329S} 
and \cite{2009A&A...505..891S}, 
and as well as the Spitzer transit photometry of
\cite{2009ApJ...699..478D}, \cite{2009ApJ...690..822K}, and 
\cite{2011A&A...526A..12D}.  
This makes this dataset an invaluable resource when
comparing the stellar activity levels from epoch to epoch.

Around the first HST visit in our data, photometric coverage was also obtained with the 40-inch telescope at Wise Observatory (Israel). 25 measurements were obtained over 45 days (JD=2455130-2455171) with an $R$ filter. This was done in order to ensure photometric coverage in case of bad weather at either site, and to verify the photometric accuracy of the monitoring. 

%

\section{ANALYSIS}
\begin{figure}
\includegraphics[width=0.475\textwidth]{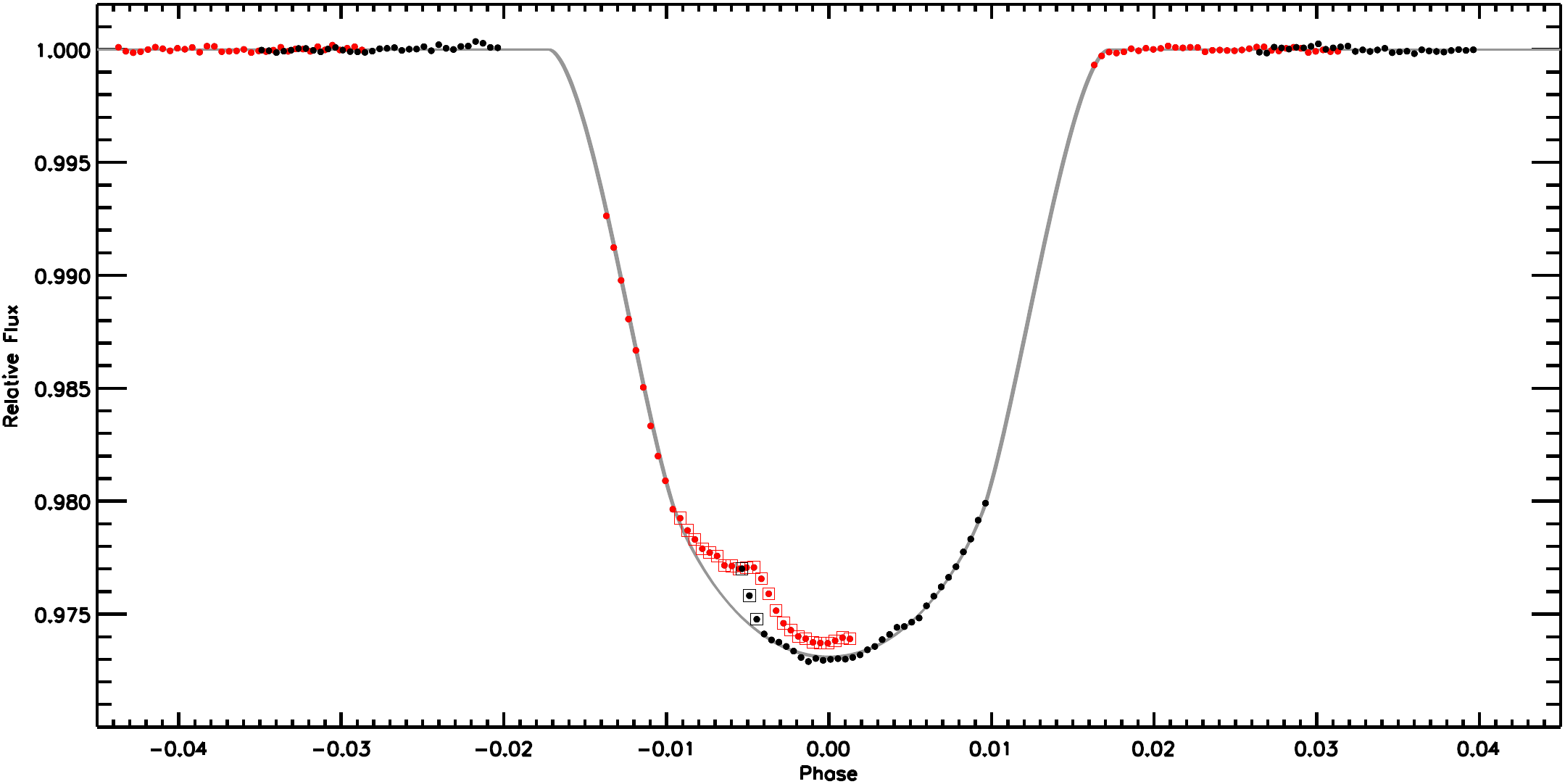}

\caption{STIS white light curve for visit 1 (black) and
  visit 2 (red) with the instrument trends removed.  The points
  showing occulted spot features are indicated with boxes.  The best
  fit transit models for both visits are shown in grey using the
  un-spotted points.}
\label{figwhitecorr}
\end{figure}

We modeled the whitelight transit light curves with the analytical transit
models of \cite{
2002ApJ...580L.171M},  
choosing to fix the central transit time, planet-to-star radius
contrast, inclination, and stellar density while fitting for the
planet-to-star radius contrast, stellar baseline flux, and
instrument systematic trends.  The system parameters
have been very accurately measured from the high S/N transit light curves of {\it HST} and
{\it Spitzer} \citep{
2007A&A...476.1347P,
2009A&A...505..891S,
2010ApJ...721.1861A,2011A&A...526A..12D}, 
and we used those of \cite{2010ApJ...721.1861A} 
for this study, taken at the longest wavelength.
The errors on each datapoint were set to the tabulated values which
is dominated by photon noise but also includes background subtraction and readout noise.
The best fit parameters
were determined 
simultaneously with a Levenberg-Marquardt least-squares algorithm 
\citep{2009ASPC..411..251M} 
using the unbinned data.
The uncertainties for each fit parameter are rescaled to
include any measured red noise (see Sec. \ref{NoiseSection}) as well as
non-unity reduced $\chi^2$ values, taking into account any underestimated
errors in the tabulated datapoints.  We find that the tabulated per-exposure errors are accurate
at small wavelength bin sizes, but are in general an underestimation
for the larger bin sizes used in this study.
A few deviant points from each light curve were cut at the 3-$\sigma$
level in the residuals, and the in-transit sections obviously crossing
stellar spots were also initially removed in the fits.  In visit 1, the beginning
3 exposures of the in-transit orbit show an obvious spot-crossing.
In visit 2, there is a large $\sim$2-4 mmag spot-like featured
centered near phase $-$0.005 (see Fig. \ref{figwhitecorr}).

\subsection{Instrument Systematic Trends}  
As in past STIS studies, we applied orbit-to-orbit flux corrections by
fitting for a fourth-order polynomial to the photometric time series, phased on the HST orbital 
period.  The systematic trends were fit simultaneously with the
transit parameters in the fit.  Higher-order polynomial fits were not statistically justified, based upon the
Bayesian information criteria (BIC; \citealt{Schwarz1978}).  Compared to
the standard $\chi^2$, the BIC 
penalizes models with larger numbers of free parameters, giving a
useful criterion to help select between different models with
different numbers of free parameters, and helps ensure that the preferred model does not
overfit the data.  The baseline flux level of each
visit was let free to vary in time linearly, described by two fit parameters.   In addition, we found it useful to also fit for further systematic
trends which correlated with the detector position of the spectra,
as determined from a linear spectral trace in IRAF and the
dispersion-direction sub-pixel shift between spectral
exposures, measured by cross-correlation.

We found that fitting the systematic trends with a fourth-order polynomial HST orbital period
correction and linear baseline 
limited S/N values to the range of 9,000 to 10,000 (precisions levels of
0.011 to 0.01\%).  These limiting values match similar
previous pre-SM4 STIS observations of HD~209458b, which also similarly
corrected for these systematic trends
\citep{2001ApJ...553.1006B, 
2007ApJ...655..564K, 
2007Natur.445..511B,
2008ApJ...686..658S}. 
With the additional correction of position related
trends (see Fig.~\ref{figwhite}), were able to increase the
extracted S/N to values of 14,000 per image, which is $\sim$80\% of the
Poisson-limited value.  These additional free parameters in the fit are also
justified by the BIC as well as a reduced $\chi_{\nu}^2$ value.  In a fit excluding the 
position dependent systematic trends for the first STIS visit, we find a BIC value
of 277 from a fit with 73 Degrees of Freedom (DOF), 8 free
parameters, and a reduced $\chi_{\nu}^2$ of 3.31.  Including the position-related trends 
lowers the BIC value to 202 from a fit with 70 DOF, 11 free
parameters, and a reduced $\chi_{\nu}^2$ of 2.19.
In the final whitelight curve fits, the uncertainty in fitting for 
instrument-related systematic trends accounts for $\sim$20\% of
the final $R_{\rm pl}/R_{\rm star}$ error budget. 

\subsection{Limb Darkening}  
At near-UV and blue optical wavelengths, the stellar limb-darkening is
strong, and in general not well reproduced by standard 1D stellar atmospheric models.
We account for the strong near-UV and optical limb darkening following
three different prescriptions; fitting for the limb-darkening coefficients,
computing limb-darkening coefficients with 1D stellar atmospheric
models, and finally using a fully 3D time-dependent hydrodynamic stellar atmospheric model.
We computed limb-darkening coefficients for the linear law
\begin{equation}  \frac{I(\mu)}{I(1)}=1 - u(1 - \mu) \end{equation}
as well as the \cite{2000A&A...363.1081C} 
four-parameter limb-darkening law
\begin{equation}  \frac{I(\mu)}{I(1)}=1 - c_1(1 - \mu^{1/2})- c_2(1 - \mu) - c_3(1 - \mu^{3/2}) - c_4(1 - \mu^{2}). \end{equation}
For the 1D models, we followed the procedures of
\cite{2010A&A...510A..21S} 
using 1D Kurucz ATLAS models\footnote{http://kurucz.harvard.edu} and
the transmission function of the G430L grating (see Table
\ref{TableLD}).  The four-parameters law is the best representation
of the stellar model intensity distribution itself, while the linear
law is the most useful in this study when fitting for the coefficients from the
transit light curves.

\begin{figure}
 \includegraphics[width=0.475\textwidth]{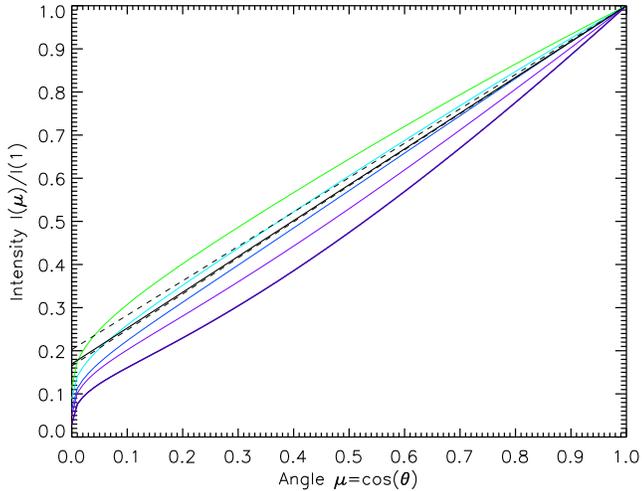}
  \caption{Stellar limb-darkening intensity profile for the STIS G430L whitelight curve
    (black - linear law) and spectral bandwidths (coloured -
    4-parameter law) for the 3D stellar
    model.  Also shown is the $\pm1-\sigma$ envelope of the best-fit linear limb-darkening coefficient from
 the whitelight transit light curve (dashed lines). }
\label{figLD}
\end{figure}

We constructed a 3D time-dependent hydrodynamical model atmosphere using
the \texttt{StaggerCode} \citep{Nordlundetal:1995} 
with a resolution of $240^{3}$ grid points, spanning
$4\mathrm{\,Mm}\times4\mathrm{\,Mm}$ on the horizontal axes and
$2.2$\,Mm on the vertical axis. The simulation has a time-average
effective temperature $\left<T_{\mathrm{eff}}\right>=5050$~K,
surface gravity $\log g=4.53$ and metallicity $\mathrm{[Fe/H]}=0.0$
\citep[based on the solar composition of][]{2005ASPC..336...25A},
which is close to the stellar parameters of
\cite{2005A&A...444L..15B}. 
Full 3D radiative transfer was computed in local thermodynamic equilibrium
(LTE) based on continuous and spectral line opacities provided by
Trampedach (2011, in prep.) and B. Plez (priv. comm.); see also
\cite{2008A&A...486..951G}.  We obtain monochromatic surface
intensities $I_{\lambda}(\mu,\phi,x,y,t)$ with a sampling of
$\lambda/\Delta\lambda=20000$ in wavelength for 17 polar angles $\mu$,
4 azimuthal angles $\phi$ and a time-series of 10 snapshots that span
$\approx30$ minutes of stellar time t, with a horizontal resolution of
120 $\times$ 120 grid points in x and y.  Limb darkening laws
$I(\mu)/I(1)$ (see Fig.~\ref{figLD}) were derived by
averaging $I_{\lambda}(\mu,\phi,x,y,t)$ over horizontal grid, azimuth
angle and time, providing a statistical representation of the surface
granulation, and integrating over each bandpass and the transmission
function; the result was normalized to the disk-centre intensity at
$\mu=1$.  A more detailed description of the 3D model will be given in
a forthcoming paper by WH. 
\begin{table*} 
\centering
 \begin{minipage}{145mm}
\caption{Linear and four-parameter limb darkening coefficients for
 HST STIS G430L}
\label{TableLD}
\begin{tabular}{llllllll}
\hline  
 Wavelength       &transit-fit                 & 1D model & 3D model &\multicolumn{4}{c}{\underline{~~~3D model four-parameter law~~~}}    \\ 
(\AA)                  &$u_{fit}$  &$u_{1D}$ & $u_{3D}$&~~~$c_1$&~~~$c_2$  &  ~~~$c_3$     &   ~~~$c_4$\\
\hline 
2900-5700         &   0.816$\pm$0.019 &  0.8326  &  0.8305 &0.5569 &-0.4209  &1.2731 &-0.4963 \\
\hline
2900-3700           & 1.01$\pm$0.03     &  0.9117    &0.9534  &0.5472& -0.9237&  1.7727& -0.4254  \\
3700-4200           & 0.91$\pm$0.01     &  0.9155    &0.9000  &0.5836& -0.8102&  1.7148& -0.5389 \\ 
4200-4700           & 0.86$\pm$0.02     &  0.8927    &0.8680  &0.5089& -0.4084& 1.3634& -0.5302  \\
4700-5200           & 0.80$\pm$0.02     &  0.7484    &0.8257  &0.5282& -0.3141&  1.1931& -0.4994 \\ 
5200-5700           & 0.72$\pm$0.02     &  0.7612    &0.7701  &0.6158& -0.3460&  1.0695& -0.4578 \\
\hline
\end{tabular}
\end{minipage}
\end{table*}

While limb-darkening is stronger at near-UV and blue
wavelengths, compared to the red and near-IR, the white-light stellar intensity
profile is also predicted to be close to linear 
(see Fig.~\ref{figLD}).  Fortunately, a linear stellar intensity profile makes
it much easier to compare fit
limb-darkening coefficients to model values, because it is less
ambiguous as to which limb-darkening law to choose and because the fit is not
complicated by degeneracies when fitting for multiple limb-darkening coefficients.  In the white light curve fits,
we performed fits allowing the linear
coefficient term to vary freely, as well as fits setting the coefficients
to their 1D and 3D predicted values with the four-parameter law.  The
resulting coefficients are given in Table \ref{TableLD}.
Given the phase coverage between our two HST visits, and the observed
occulted stellar spots in each visit, it is much more straightforward
to compare model limb-darkening from visit 1, which is largely
occulted-spot free.  
Using the linear limb-darkening law, we find $u_{fit,white}=0.816\pm0.019$
for visit 1 (see Table \ref{TableLD}).  The fit coefficient is within 1-sigma
of an appropriate ATLAS model ($u_{1D model}= 0.8326$ for T$_{\mathrm{eff}}$=5000~K, log
g=4.5, [Fe/H]=0.0, and $v_{turb}=2 km/s$) 
and within  1-sigma of the 3D model, which predicts $u_{3D}= 0.8305$.  
For visit 1, the freely fit linear limb-darkening model
gives a $\chi^2$ value of 265 for 78 Degrees of Freedom (DOF) giving a
BIC value of 315.
We find that the fit is significantly worse when adopting the 1D ATLAS models with the
four-parameter law, with a $\chi^2$ value of 286 for 79 DOF and a BIC value of
331.  Using the coefficients derived from the 3D time-dependent hydrodynamical
model provides an equally good fit as the linearly-fit model, with a
$\chi^2$ value of 27 for 79 DOF and a BIC of 315.  The improvement
from the 1D model to the 3D model is clear, with the 3D model
providing a much better fit to the transit light curve.  Using the BIC
as a guide, the 3D model also performs just as well as the freely fit
linear limb-darkening law.



\begin{table} 
\centering
\begin{minipage}{80mm}
\caption{Visit 1 transit light curve fits.}
\label{TableRad}
\begin{tabular}{lllllll}
\hline
Wavelength&  \multicolumn{3}{c}{\underline{fit limb-darkening}}& \multicolumn{3}{c}{\underline{3D limb-darkening}}\\
(\AA) & $\chi^2$ & DOF & BIC & $\chi^2$ & DOF & BIC \\
\hline
2900-3700       &  208 &  83 &   258   &  201 & 84  & 247 \\
3700-4200       &  116   &  82 &    166  & 115 &  83 & 160 \\
4200-4700       &  132 & 82  &   182  &  132 & 83  & 177 \\
4700-5200       &  161 & 83  &  211  & 162 & 84  & 207 \\
5200-5700        & 114   &  82 &  164   & 119 & 83  &  164\\
\hline
\end{tabular}
\end{minipage}
\end{table}

As a further test, we also tried fitting a quadratic
limb-darkening law, fitting for both coefficients ($u_+$ and $u_-$ in a minimally
correlated manner) but found no significant improvement, as the quadratic term
was not particularly well constrained from our observations.
%
\begin{figure*}
 \includegraphics[width=14cm]{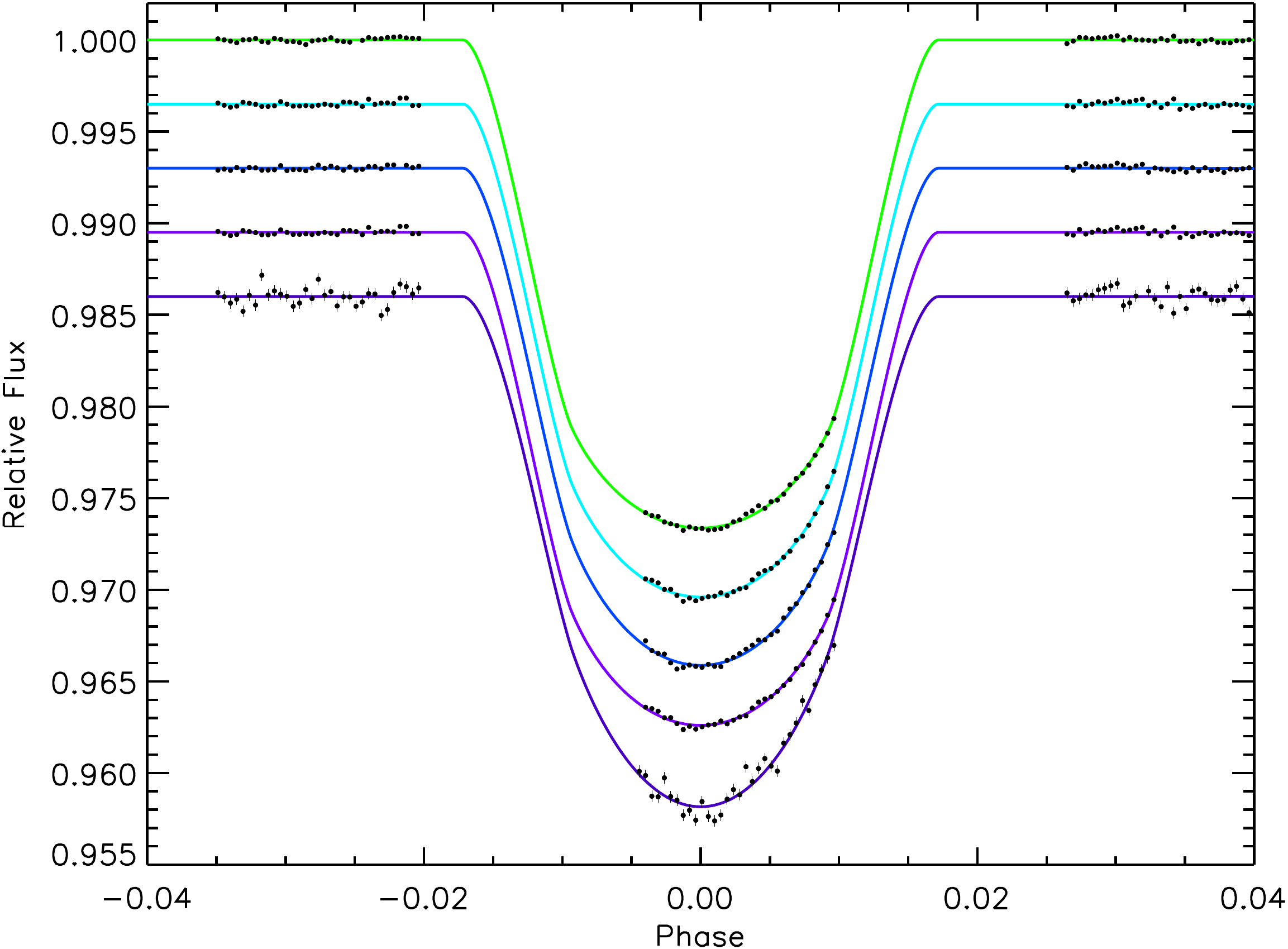}
  \caption{STIS G430L visit1 lightcurves at (bottom to top)
    3300, 3950, 4450, 4950, and 5450~\AA\ used to derive
    the planetary radius. The lightcurves have been corrected for
    instrumental effects and the first few exposures of the transit affected by
    occulted starspots have been removed.
    An arbitrary flux offset has been applied and instrumental trends removed for clarity.  The best-fit transit light curve models are
    also overplotted.}
\label{figspectra}
\end{figure*}
\begin{figure}
 \includegraphics[width=0.47\textwidth]{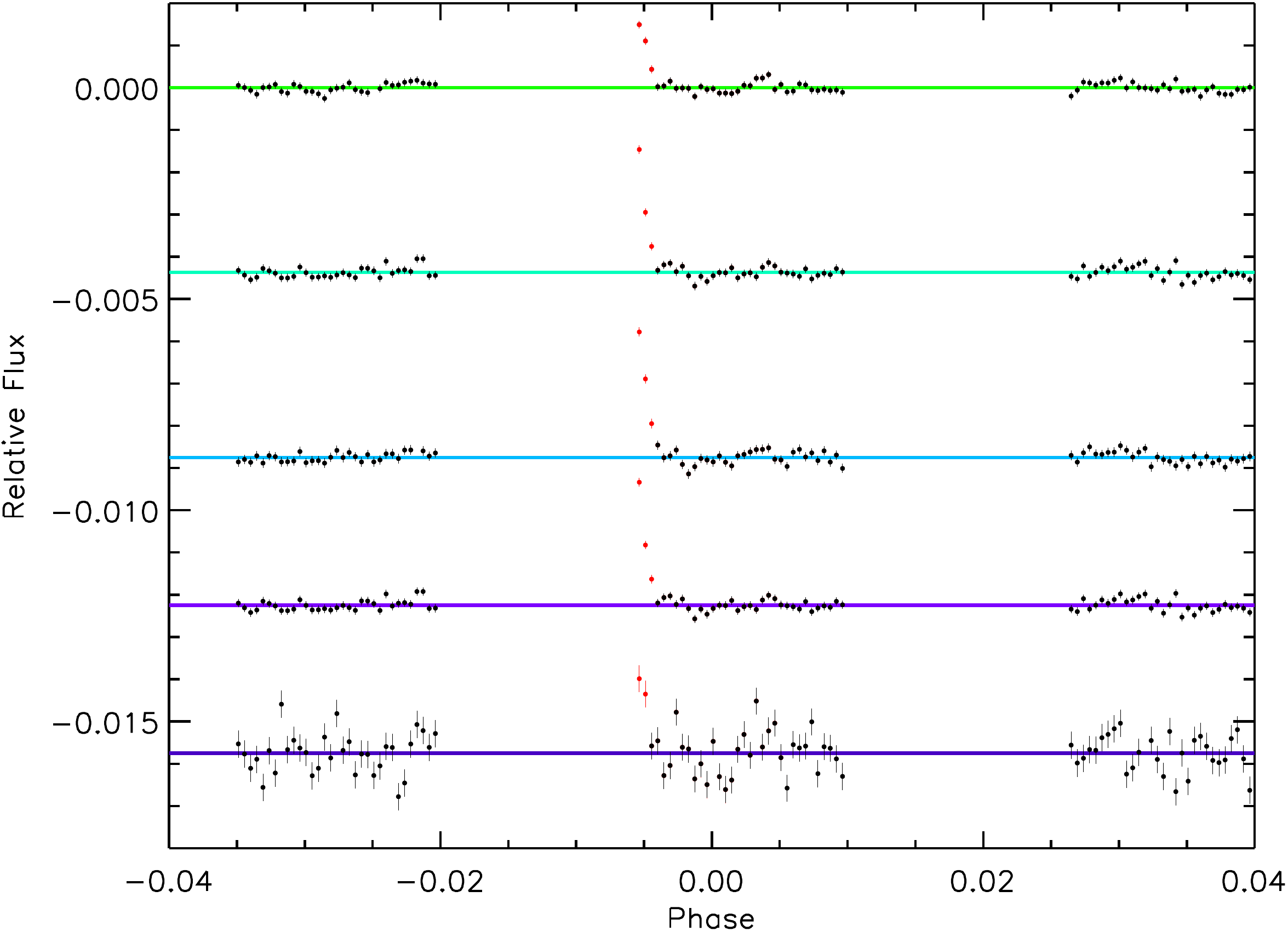}
  \caption{STIS G430L visit1 lightcurve residuals at (bottom to top)
    3300, 3950, 4450, 4950, and 5450~\AA\ with an arbitrary offset.  Occulted-spot
    features not used in the fit are shown in red.}
\label{figspectraresid}
\end{figure}

\subsection{Transmission Spectrum Fits}
\label{specfits}
To measure the broad transmission spectral features and compare them with the previous
ACS results, we also extracted 500~\AA\ wide spectral bins from the
spectral time series (see Fig.~\ref{figspectra} and Fig.~\ref{figspectraresid}).  The resulting five
bands were analyzed and individually fit in the same manner as the
white light curve fits.  Comparing transit fits using freely fit
limb-darkening coefficients to those set of the 3D model, significant overall improvement is seen with the 3D
model (see Table \ref{TableRad}).  As in the whitelight curve tests, the 3D model also handily
outperforms fits using the 1D ATLAS models.  At the shortest
wavelength bins, adopting the 3D model limb-darkening gives a lower $\chi^2$ value, even
with less free parameters in the transit lightcurve fit.  Given the
overall excellent performance of the 3D model, we adopt the predicted
four-parameter 3D model coefficients for our final light curve fits.
Comparing the resulting planetary radii values between those using fit
limb-darkening and the 3D model, we find similar radii values for the
middle two wavelength bins starting at 4200 and 4700 \AA.  The bluer two wavelength bins give
increasingly larger radii values when using the 3D model.  This trend
is repeated when using the 1D ATLAS model as well, though exaggerated
in the 1D case.
The better fits using the 3D models and expectation that the
limb-darkening at those wavelengths is significantly non-linear, indicates that larger planetary
radii values toward the blue are highly likely, though some caution is in
order, as these are the very first results using limb-darkening from a
3D stellar model.  Such models will no doubt need further vetting at other wavelengths
and against other datasets before firm confidence is ascertained.  However, the overall improvement of the stellar
limb-darkening from the 1D model to the 3D model is very encouraging,
and very much in line with existing tests
comparing such models to the Sun (see \citealt{2009ARA&A..47..481A} and
\citealt{2008ApJ...686..658S}).  There is a broad overall trend of 1D
models over predicting the strength of limb-darkening
compared to 3D models and solar data, due primarily to a steeper atmospheric temperature
gradient in the 1D case. 
In 3D solar model atmospheres, convective motions lead to a shallower
mean temperature profile near the surface and therefore to an overall
slightly weaker predicted limb-darkening.
The resulting planetary radius ratios from visit 1 are given in
Table~\ref{TableAveRad}, which use the 3D model limb-darkening 
four-parameter law coefficients of Table~\ref{TableLD}. 
A similar transmission spectral shape is also obtained if the 1D
model limb-darkening is used.

The large spot feature in visit 2 appears to affect most of the measurements between second and third contact, making it
less straightforward to measure the planetary radii.
We find that the planetary radius values of visit 2 are all in
agreement with those from visit 1, when fitting for the visit 2 radii using the unaffected datapoints during
ingress and a few points after
second contact,  fixing the limb-darkening
to the 3D model coefficients (see Fig. \ref{figwhitecorr}).  
In Section \ref{spots} we describe a
procedure used for visit 2 to fit for the occulted-spot and determine
the wavelength-dependent planetary radii.

\subsection{Noise Estimate}
\label{NoiseSection}
We checked for the presence of systematic errors correlated in time (``red noise'') 
by checking whether the binned residuals followed a $N^{-1/2}$ relation when
  binning in time by N points.  In the presence of red noise, the
  variance can be modeled to follow a $\sigma^2=\sigma_w^2/N+\sigma_r^2$ relation,
  where $\sigma_w$ is the uncorrelated white noise component, while
  $\sigma_r$ characterizes the red noise \citep{
2006MNRAS.373..231P}.  
We found no significant 
  evidence for red noise in the highest S/N lightcurves of visit 2 when binning on timescales up to
  15 minutes (10 exposures).  In visit 1 we found low levels of
  residual systematic errors, with $\sigma_r$ values
  around 2.5$\times10^{-5}$, which corresponds to a limiting S/N level of
  $\sim$40,000.  We subsequently rescaled the photometric
  errors to incorporate the measured values of $\sigma_r$ in the 1$\sigma$ error bars of the
  whitelight and spectral bin fits.  The remaining low levels of systematic
  trends in visit 1 do not have a significant effect on our final
  transmission spectrum.  A limiting precision of
  2.5$\times10^{-5}$ translates into a limiting value of 8$\times10^{-4}$
  $R_{\rm pl}/R_{\rm star}$, which is $\sim$5 times smaller than the atmospheric
  scale height at 1340~K.

\subsection{CORRECTING FOR THE STELLAR ACTIVITY}
\label{spots}

A transmission spectrum for a transiting planet with an active host star can be affected by the
presence of stellar spots, both by spots occulted by the planet and by non-occulted
spots visible during the epoch of the transit observations \citep{
2008MNRAS.385..109P, 
2010ApJ...721.1861A}. 
When the planet masks a starspot on the surface of the star during a transit (``occulted spot''), the flux rises proportionally to the dimming effect of the spot on the total stellar flux. The presence of a starspot in a region of the stellar disc not crossed by the planet (``unocculted spot'') causes a difference in surface brightness between the masked and visible parts of the stellar disc, slightly modifying the relation between transit depth and planet-to-star radius ratio. In both cases the effect is wavelength-dependent and must be corrected to obtain the planetary transmission spectrum.

Fortunately, in our observations these effects can be adequately corrected. The lightcurves are precise enough for the spot crossing events to be clearly identified, while for reasonable assumptions on the amount of spots on HD 189733, the influence of unocculted spots on the transmission spectrum is small enough that a first-order correction is possible, using the measured flux variations of the star during the HST visits and the temperature contrast of the spots inferred from occulted spots, as already discussed in  \citet{2008MNRAS.385..109P}. We describe in the following paragraphs the treatment of occulted and unocculted starspots in our analysis.

\subsubsection{Occulted spots}
\label{occulted}
The STIS data shows clear evidence for two spot occultations by the planet, one in each visit. The flux rise is stronger at bluer wavelengths, and the timescale of the events corresponds to the motion of the planet across features on the star, so that the interpretation in terms of a masked cooler feature on the surface of the star is natural. The first event shows only the end of an occultation, while the second shows a full occultation and can be analyzed in detail. 

\begin{figure*}
\includegraphics[width=14cm]{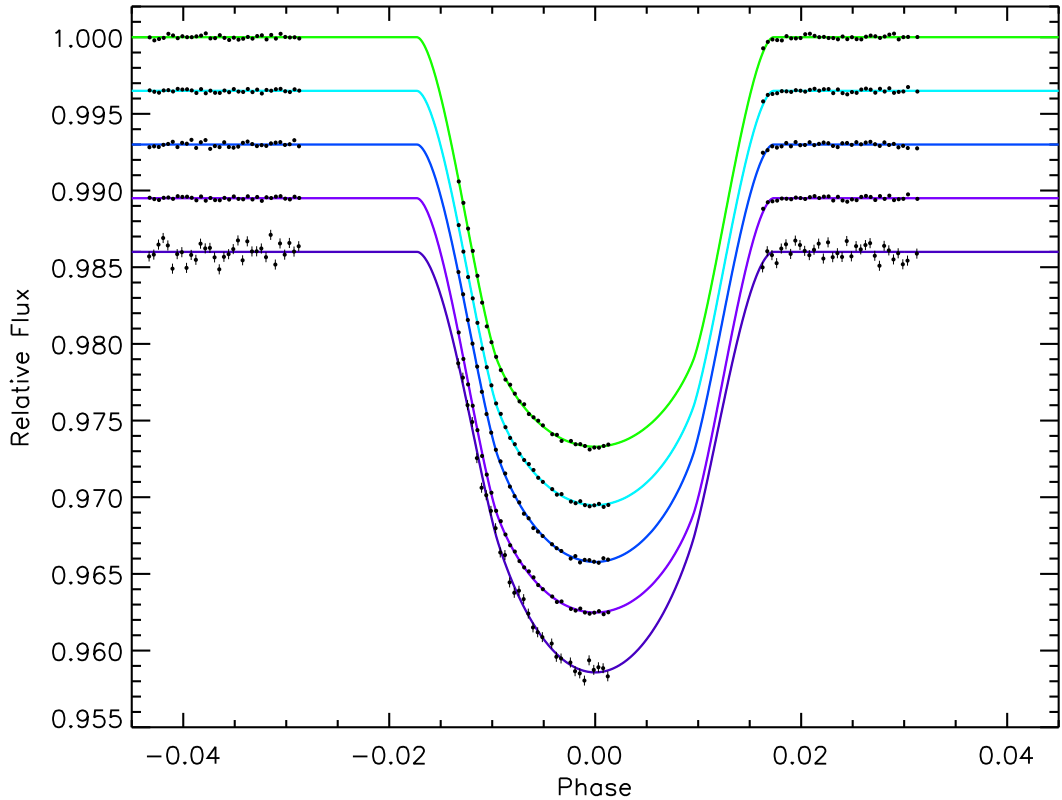}
 \caption{Same as Fig. \ref{figspectra} for visit2.  For clarity, the
   occulted spot has been corrected from the lightcurves and is shown in Fig. \ref{OccultedSpot}.}
\label{figspectraV2}
\end{figure*}

\begin{figure}
\includegraphics[width=0.475\textwidth]{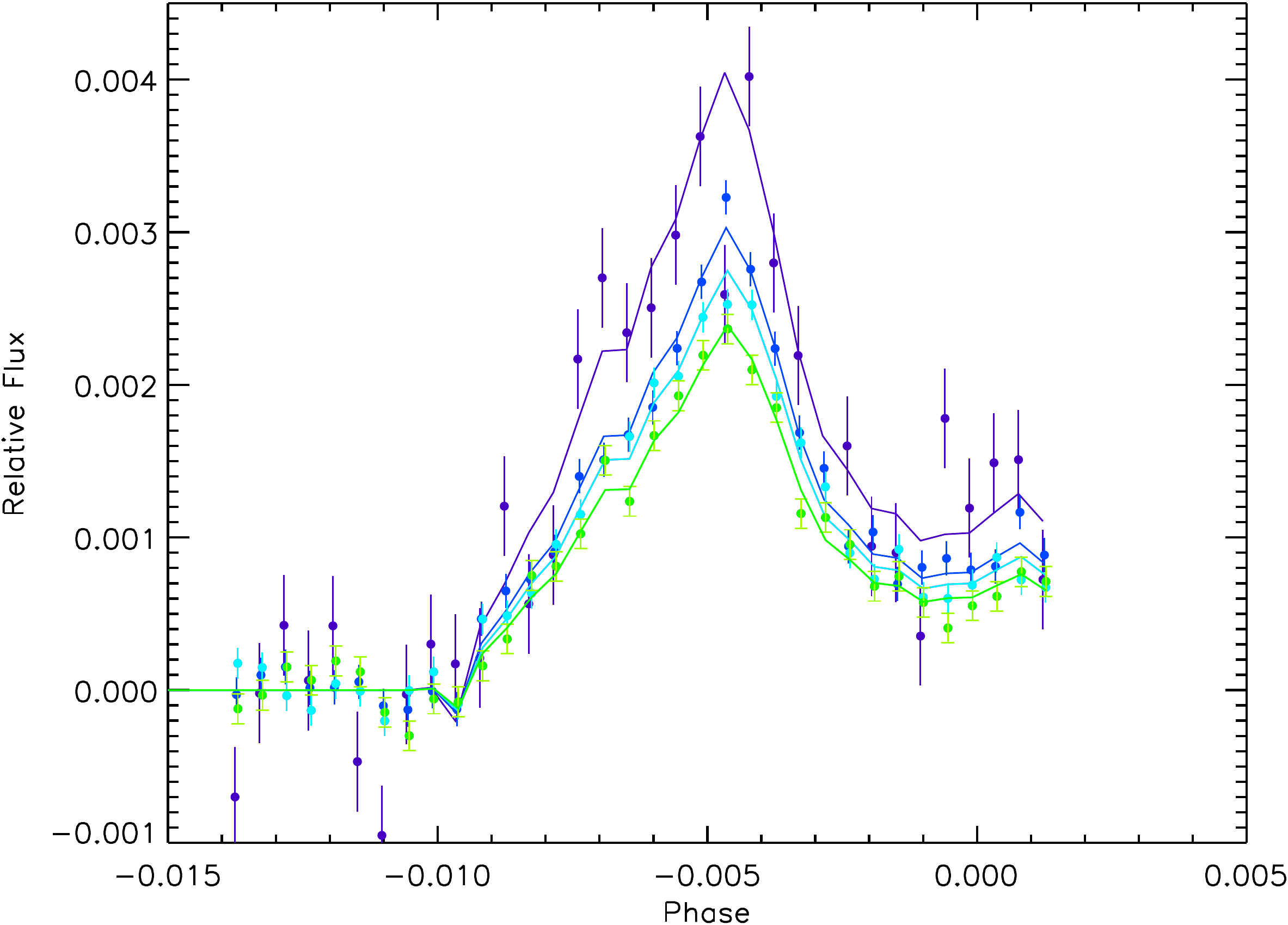}
\caption{Plotted is the occulted spot feature from visit 2 at (top to bottom)
   3300, 3950, 4450, 4950, and 5450~\AA\, along with
 the bestfit spot solution.}
\label{OccultedSpot}
\end{figure}

\begin{figure}
\includegraphics[width=0.475\textwidth]{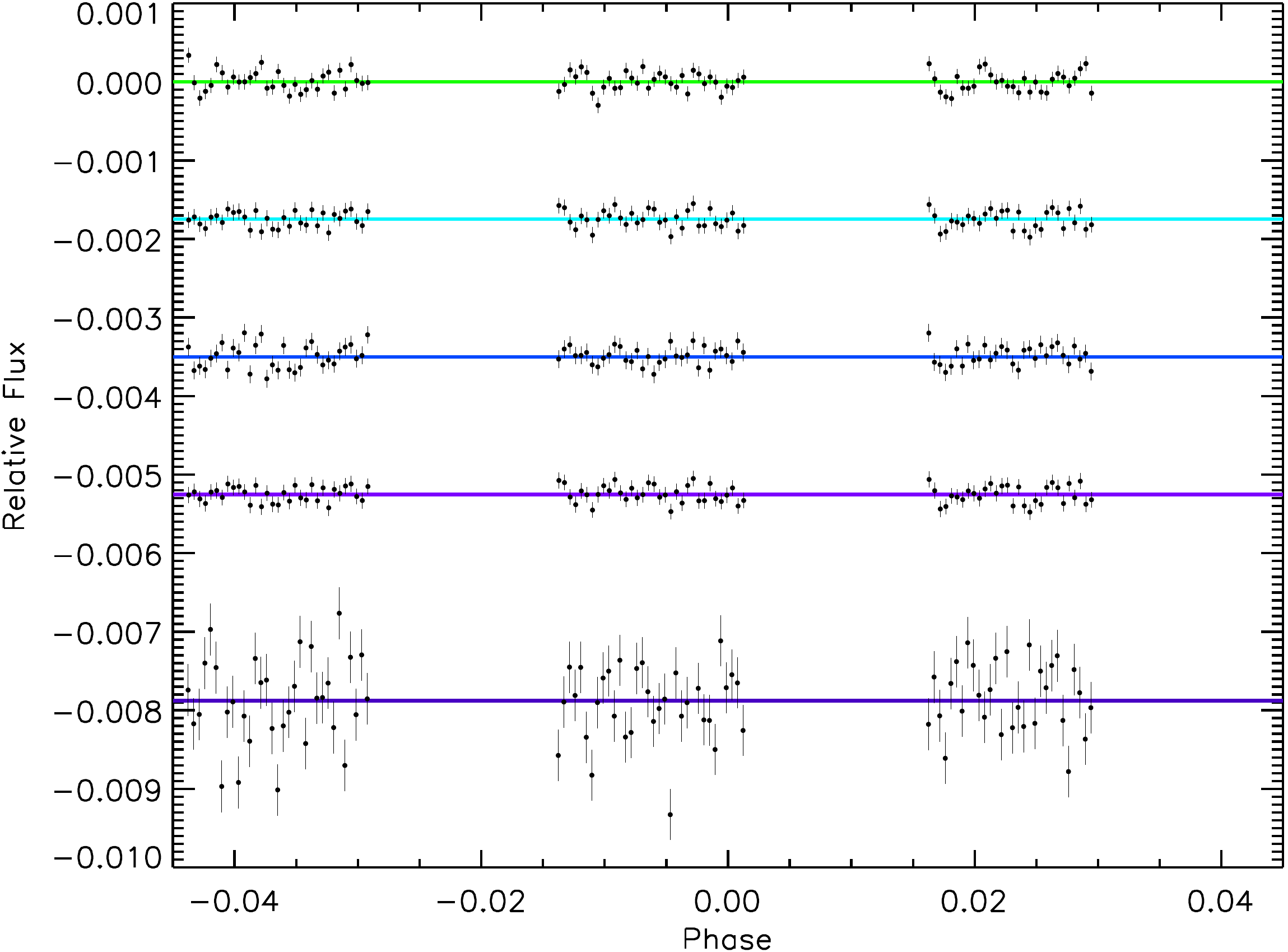}
 \caption{STIS G430L visit2 lightcurve residuals at (bottom to top)
   3300, 3950, 4450, 4950, and 5450~\AA\ with an arbitrary offset.}
\label{figspectraVis2residB}
\end{figure}

Previous studies of the variability of HD 189733, such as the
space-based lightcurve with the MOST telescope
\citep{2007ApJ...671.2129C} 
and intensive spectroscopic monitoring with the SOPHIE spectrograph
\citep{2009A&A...495..959B} 
show that the variability of the star can be well
modeled both in flux and spectrum with cooler starspots modulated by
stellar rotation. We can test this hypothesis by computing the
wavelength dependence of the amplitude of the flux rise caused by the
spot occultation in the second visit. For this purpose, we have fitted
simultaneously the planetary transit (see Fig.~\ref{figspectraV2}) with a wavelength-depend radius
ratio and the effect of the spot with a wavelength-dependent factor
for the amplitude of the spot feature. We take the shape of the spot
feature from the lightcurve integrated over all wavelengths, but let
it vary by a free multiplying parameter in each of five 500 \AA\
wavelength bands. The results are shown in Figs.~\ref{OccultedSpot}, \ref{figspectraVis2residB}, and ~\ref{Figspotcorr}. We
then compare these amplification factors to expectations for the
occultation of a cooler area on a $T_{\rm eff}=5000$ K star. We assume
that both the star and the spot have spectral energy distributions
well described by Kurucz atmosphere models, and use spot temperatures
from 3500 K to 4750 K. Figure~\ref{Figspotcorr} compares the observed
amplification factors to the model expectations. In that case the flux
rise $\Delta f_\lambda$ at wavelength $\lambda$ compared to a reference
wavelength $\lambda_0$ will be  
\begin{equation}
\Delta f_\lambda / \Delta f_{\lambda_0} = \left(1-\frac{F^{T_{\rm spot}}_\lambda}{F^{T_{\rm star}}_\lambda}\right)/\left(1-\frac{F^{T_{\rm spot}}_{\lambda_0}}{F^{T_{\rm star}}_{\lambda_0}}\right)
\label{eq1}
\label{EQradiflux}
\end{equation}
where $F^T_\lambda$ is the surface brightness of the stellar atmosphere models at temperature $T$ and wavelength $\lambda$.
We also plot on the figure the results of the same procedure for the ACS visit in Pont et al. (2008). 

Figure~\ref{Figspotcorr} shows that the spectrum of the occulted spots is well constrained, corresponding to the models with 
spots at least 750~K cooler than the stellar surface.  
Although the difference spectrum of the occulted spot is coherent with
the expectation in overall shape, the 5000 \AA \ MgH feature is weaker than
expected. This does not have a significant impact on the present
study, but we point it out as a possible intriguing feature of
starspots on HD~189733. 



\subsubsection{Unocculted spots}

Since the timescale of spot variability is much longer than the planet crossing time during transits ($P_{\rm rot}\simeq 12 $ days vs $T_{\rm tr}\simeq 1.5$  hours for HD 189733), unocculted spots can be considered as stationary during a transit, and their effect will correspond to a fixed wavelength-dependent correction on the transit depth. 

\begin{figure}
%
 \includegraphics[width=0.475\textwidth]{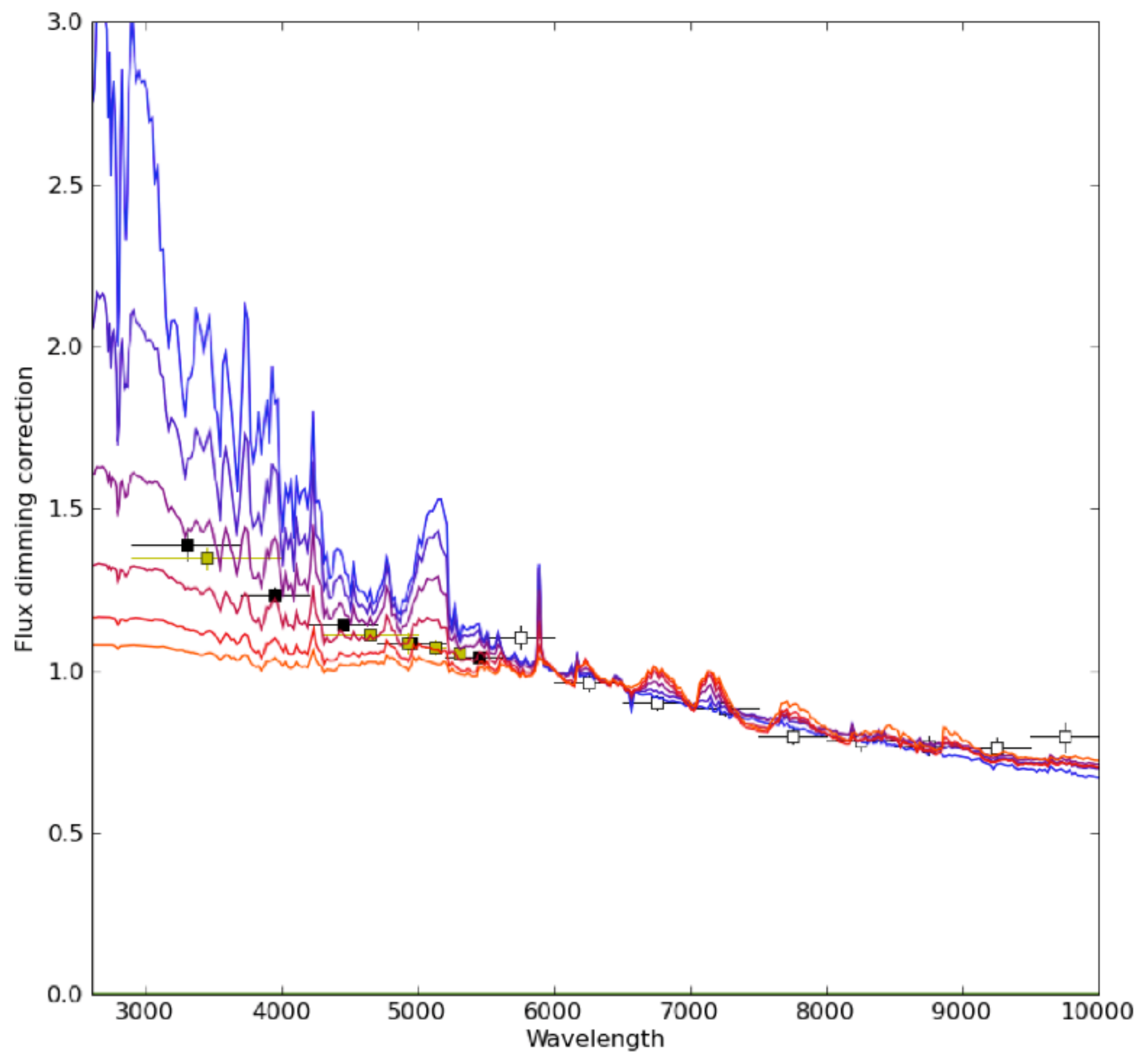}
\caption{Spectral signature of the stellar spots occultation derived from the
   STIS G430L (closed black and green symbols) and ACS (open symbols)
   data.  The
 spot is modeled with stellar atmospheric models of different
 temperatures ranging from 4750 to 3500~K in 250~K intervals (blue
 to orange respectively), and $T_{\rm eff}=5000$~K for the stellar
 temperature.}
\label{Figspotcorr}
\end{figure}

As in Pont et al. (2008), we model the effects of unocculted spots by assuming
that the emission spectrum of spots corresponds to a stellar spectrum
of lower temperature than the rest of the star covering a fraction of
the stellar surface and assuming no change in the surface brightness outside spots.  
We neglect the effect of
faculae on the transmission spectrum. 
The spots then lead to an overall dimming of the star.  Under these
assumptions, to reach the same level of flux dimming, higher spot
temperatures require a greater fraction of stellar surface covered by
spots.
These assumptions are now supported
by the behaviour in several HST visits. The signature on the flux of
occulted spots is frequently seen (to the point in fact that no
entirely spot-free visit was encountered among our 9 visits), and the
occulted spots observed have the expected red signature \citep[Fig.~\ref
{Figspotcorr};][]{2008MNRAS.385..109P}.  No detectable facula occultation is observed (a
facular occultation would result in a sharp flux drop during the
transit with a blue spectral signature).  In Section~\ref{SecERAC} below, we present a further test in support of the validity of this assumption.

\begin{figure*}
\includegraphics[width=0.70\textwidth,angle=0]{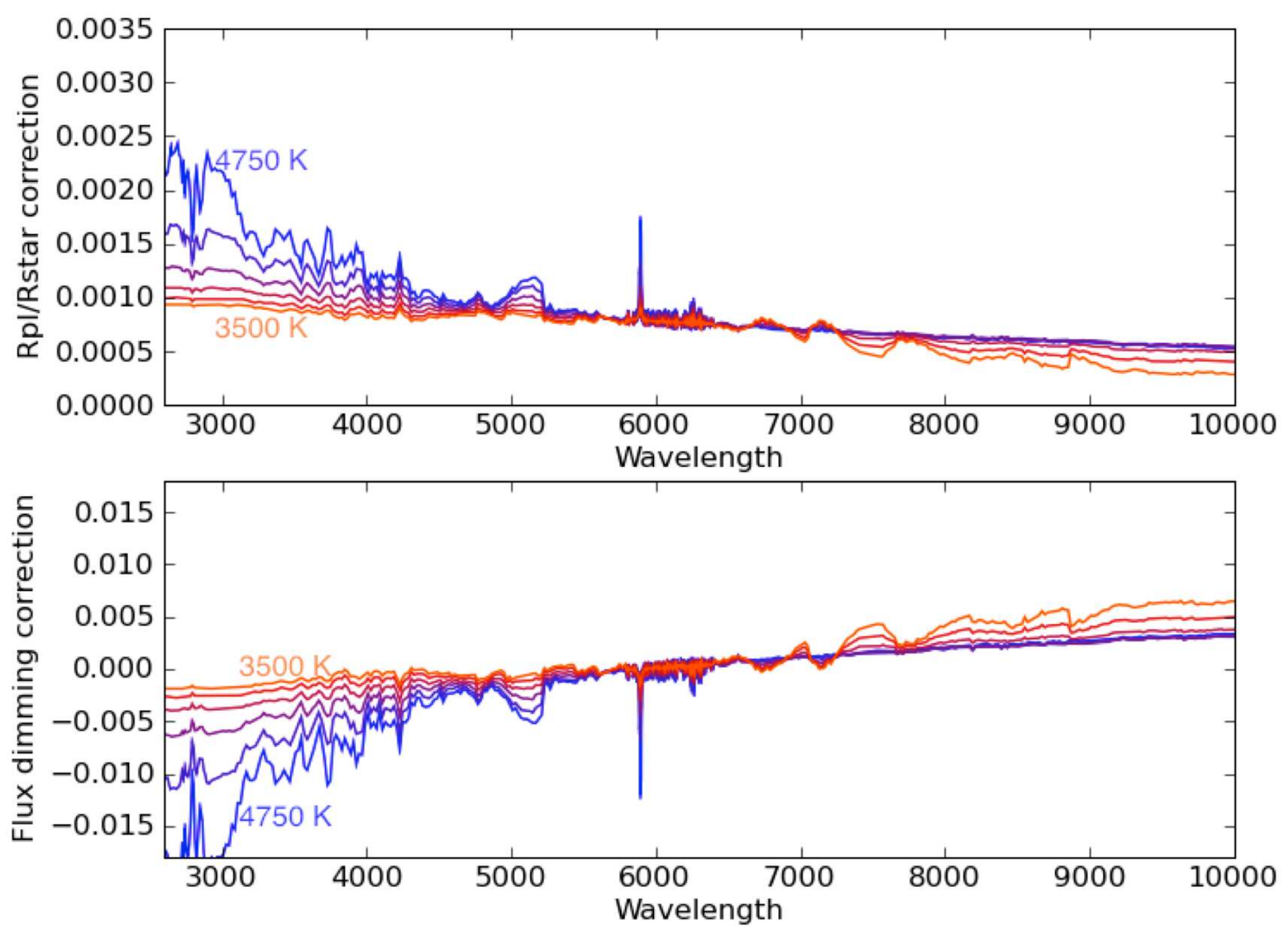}
\caption{(Top) The effect of spot coverage on the transmission
 spectrum and a 1\% stellar flux drop at 6000~\AA\ where the radius ratio changes by $\Delta R_{pl}/R_{star}=0.0008$.  The
 spot is modeled with stellar atmospheric models of different
 temperatures ranging from 4750 to 3500~K in 250~K intervals (blue
 to orange respectively), and $T_{\rm eff}=5000$~K for the stellar
 temperature. (Bottom) The relative stellar flux, $\Delta f_{\lambda}$, for the different spot
 temperatures compared to the flux at 6000~\AA, assuming a 1\% drop in stellar flux at that wavelength.}
\label{spotcorr}
\end{figure*}

Under the assumption that the stellar flux is a combination of a
surface at $T=T_{\rm star}$ and spots at $T=T_{\rm spot}$ causing a
total dimming $\Delta f(\lambda_0, t)$, the corrections to the transit depth
$d$ at wavelength $\lambda$ and
radius ratio due to unocculted spots will be: 
\begin{equation}
\frac{\Delta d} {d} = \Delta f(\lambda_0, t) \left( 1- \frac{F^{T_{\rm spot}}_\lambda}{F^{T_{\rm star}}_\lambda}\right)/\left(1- \frac{F^{T_{\rm spot}}_{\lambda_0}}{F^{T_{\rm star}}_{\lambda_0}}\right)
\label{eqradspotrelationd}
\end{equation}
and
\begin{equation}
\Delta (R_{\rm pl}/R_{\rm star}) \simeq \frac{1}{2} \frac{\Delta d}{d}
(R_{\rm pl}/R_{\rm star}).
\label{eqradspotrelation}
\end{equation}
A similar formalism is given in \cite{2010arXiv1012.0518B}. 
Figure~\ref{spotcorr} shows the correction for unocculted spots for
$\Delta f(\lambda_0, t)$=1\% at $\lambda_0=6000$ \AA\ for different spot
temperatures. The influence over the transmission spectrum on the STIS
wavelength range is of the order of $2 \times 10^{-3}$ $R_{\rm
  pl}/R_{\rm star}$, which is $\sim$5 times smaller than the observed
variations (see Section~\ref{discussion}), suggesting that the
uncertainties on the first-order correction for unocculted spots only
add a small contribution to the final errors on the planetary
transmission spectrum (Pont et al. 2008). 

To apply the correction for unocculted spot, we need the following quantities:

\begin{enumerate}
\item{the variation in spot dimming $\Delta f(\lambda_0,t)$ between the different HST visits}
\item{an estimate of the absolute level of the stellar flux
    corresponding to a spot-free surface $ f(\lambda_0, t_{\rm ACS})$ }
\item{an estimate of the effective temperature of the spots $T_{\rm spots}$}
\end{enumerate}

\label{effectspots}

\subsubsection*{(i)Visit-to-visit flux variations}

We analyzed the ground-based photometric data with a spotted star model using the
same approach as in \cite{2010arXiv1008.3859P}, 
described in Aigrain et al. (2011).  We generate a large number of evolving spots modulated by
the stellar rotation, and instead of finding a unique solution that
reproduces the observed photometry, we explore the space of all
possible solutions. The scatter between different solutions using
different number of spots and different spot parameters is used as an
indicator of the uncertainty on the inferred stellar flux at a given time. 

HD 189733 is now a very well-studied star, and as a result a number of
key parameters for the spot simulations can be set to known values,
namely the stellar size ($R=0.755$ R$_\odot$, Pont et al. 2007),
rotation period ($P_{\rm rot}$=11.9 days,
\citealt{2008AJ....135...68H}),  
and spin-orbit angle (negligible, \citealt{2006ApJ...653L..69W}).   
Thanks to this and the excellent coverage of the photometric monitoring around the time
of the HST visits, the light curve interpolation is very stable and
the uncertainty on the amount of dimming by startspots at the time of
the STIS measurements is not more than $2 \times 10^{-3}$ (see Fig.~\ref{figwise}).  

Figure~\ref{figwise} gives five realizations of our many-spot model
around the time of the HST visits. The number of spots used
in the solution is 6, 12, 24, 48 and 96. Both the Wise and APT data
are shown on the same plot, although they are taken in different
filters and the amplitude of variability will be slightly different.   All the
solutions give very similar estimates for the brightness of the star
at the moment of the first STIS/HST visit.  They differ in less
well-constrained part of the lightcurve, but this does not impact the
present analysis. The flux in the figure is normalized to the flux at
the time of the third ACS visit, used as baseline in Pont et
al. (2007).  Fortunately the APT coverage extends across both periods,
allowing a direct comparison.  

The rectangles on Fig.~\ref{figwise} show the values we adopted for the flux level during the STIS/HST visits. For the first visit, we use the APT level since the ($b+y$)/2 filter is near the central wavelength of the STIS measurements, while the $R$ filter used in the Wise measurements is redder. 
Section \ref{SecERAC} provides an independent check of the validity of the
spot-coverage estimates from the photometric monitoring, showing 
that the variations in observed transit depths are in close agreements
with the expected variations.


\subsubsection*{(ii) Flux level of spot-free surface}

\label{constant}
Pont et al. (2008) estimated the dimming level compared to a spot-free surface for the ACS visit at 1\%.  This component of the spot
coverage is not measured by the photometric monitoring and must be inferred from the variability lightcurve.
The long-term photometric 
monitoring with the APT over several years suggests that this value is realistic, since year-to-year variations of the maximum flux do not exceed the percent level.
Pont et
al. (2008; see their Fig. 2, lower-right panel) computed the effect
of such a constant spot background, and the same estimates are valid
for the present papers. The effect is a 200-400 km difference in transit
radius (2-4$\times 10^{-3}$ on the radius ratio) per percent of stellar light blocked by the additional spot
coverage  over the 6000-10000 \AA\ wavelength range. Thus, a large amount of background spot coverage would lower
the blueward slope of the measured transmission spectrum.  

The statistics of spot crossings on our HST data as a whole (3 visits
with ACS, 2 with STIS/G430L and 4 with STIS/G750M) suggest that the
global spot coverage of the star correspond to expectations from the
photometric variability and do not point to a large additional spot
background. The statistics of occulted starspots during the 9 visits
corresponds to a flux drop of 2.8$\pm$0.8 \% at $\lambda_0$ when extrapolated to the
entire star surface.  This of course assumes that the planet crossings sample a
representative part of the star.  The planetary transits cover the
$29-55^o$ latitude interval. On the Sun, it is wellknown that spots
cluster at low latitude. However, spots for more active stars are
found to be distributed at higher latitudes than on the Sun, and cover
a larger latitude range (e.g. \citealt{2007ApJ...659L.157B}).


Nevertheless, when considering the transmission spectrum of HD 189733
over a wider wavelength range - for instance when comparing the
visible effective radius with the values observed in the mid-infrared
with Spitzer (e.g. \citealt{2010ApJ...721.1861A}) 
this source of uncertainty should be considered.

\subsubsection*{(iii) Spot temperature}

Pont et al. (2008) found that the spectral signature of the largest
spot feature occulted during the ACS visit was well described by
assigning $T\sim 4000$ K to the spot. In Section~\ref{occulted} above
we found $T\sim 4250$ K for the spot occulted during our second STIS
visit. Fig.~\ref{Figspotcorr} shows that spot temperatures in this
range describe the effect of both occulted spots
adequately. Spot-to-star temperature differences in this range are
compatible with what is measured for solar spots.

\subsubsection{Correcting for unocculted spots}

We used Kurucz stellar atmosphere models to compute the unocculted spot
corrections. Following the discussion above, we use $T_{\rm spot}=4250 \pm 250$ K and
a spot-free flux level 1.0-2.8~\% above the ACS visit flux. We apply this correction to compute the 
planetary atmosphere transmission spectrum and propagate the uncertainties in the correction.
Figure~\ref{spotcorr} shows the effect of unocculted starspots
amounting to a 1-percent dimming at 6000~\AA\ on the transmission
spectrum, and Table~\ref{deltas} gives the corrections as a fraction of transit depth integrated over 500 \AA\ passbands.  Overall, the effect is very small, but not negligible in the
context of the highly accurate HST transmission spectroscopy. 

\begin{table} 
\centering
 \begin{minipage}{80mm}
\caption{Fit STIS G430L Radius Ratios in five wavelength bands and spot
  corrected values using Table \ref{deltas} and Eq. \ref{eqradspotrelation}
  with $\Delta f(\lambda_0, {\rm visit~1})$=0.017 and $\Delta  f(\lambda_0, {\rm visit~2})$=0.013.}
\label{TableAveRad}
\begin{tabular}{llll}
\hline  
$\lambda$        &Visit 1&Correction & Visit 1 Corrected\\
(\AA)                   & $R_{pl}/R_{star}$&$\Delta R_{pl}/R_{star}$&$R_{pl}/R_{star}$\\
\hline 
3300           &0.15985$\pm$0.00042 &-0.00172 &0.15813$\pm$0.00048 \\
3950           &0.15872$\pm$0.00019 &-0.00161 &0.15711$\pm$0.00019 \\ 
4450           &0.15803$\pm$0.00023 &-0.00150 &0.15653$\pm$0.00023  \\
4950           &0.15770$\pm$0.00022 &-0.00154 &0.15616$\pm$0.00022  \\
5450           &0.15744$\pm$0.00022 &-0.00138 &0.15606$\pm$0.00022 \\
\hline
\hline  
$\lambda$        &Visit 2&Correction & Visit 2 Corrected\\
(\AA)                   & $R_{pl}/R_{star}$&$\Delta R_{pl}/R_{star}$&$R_{pl}/R_{star}$\\
\hline 
3300           &0.15803$\pm$0.00050 & -0.00132&0.15671$\pm$0.00056\\
3950           &0.15796$\pm$0.00024 & -0.00123&0.15673$\pm$0.00024\\
4450           &0.15787$\pm$0.00014 & -0.00144&0.15673$\pm$0.00014\\
4950           &0.15765$\pm$0.00011 & -0.00117&0.15648$\pm$0.00011\\
5450           &0.15725$\pm$0.00012 & -0.00105&0.15620$\pm$0.00012\\
\hline
\hline  
\end{tabular}
\begin{tabular}{ll}
$\lambda$        &Combined\\%
(\AA)                   & $R_{pl}/R_{star}$\\
\hline 
3300           &0.15754$\pm$0.00042 \\
3950           &0.15696$\pm$0.00015 \\
4450           &0.15667$\pm$0.00012 \\
4950           &0.15641$\pm$0.00010 \\
5450           &0.15617$\pm$0.00011 \\
\hline
\end{tabular}
\end{minipage}
\end{table}
\begin{table}
\begin{minipage}{85mm}
\caption{Unocculted spot corrections to the transmission
 spectrum with spot temperature T${\rm spot}= 4250 \pm 250$ causing
1\% dimming at 6000~\AA.  On this scale, the absolute value of the
shift is 1\% of the transit depth at 6000~\AA\ by definition.}
\begin{tabular}{ll}
\hline
Passband & Spot correction \\
(\AA) & $\Delta d / d$ [$\times 10^{-3}$]\\
\hline
3200-3750 & 2.92$\pm$1.13\\
3750-4000 & 2.08$\pm$0.94\\
4250-4750 & 1.21$\pm$0.45\\
4750-5250 & 1.51$\pm$0.53\\
5250-5750 & 0.31$\pm$0.17\\
5750-6250 & ~~0.0$\pm$0.0\\
\hline\end{tabular}
\label{deltas}
\end{minipage}
\end{table}

\subsubsection{Empirical Radius-Activity Level Correlation}
\label{SecERAC}

\begin{figure}
 \includegraphics[width=0.475\textwidth]{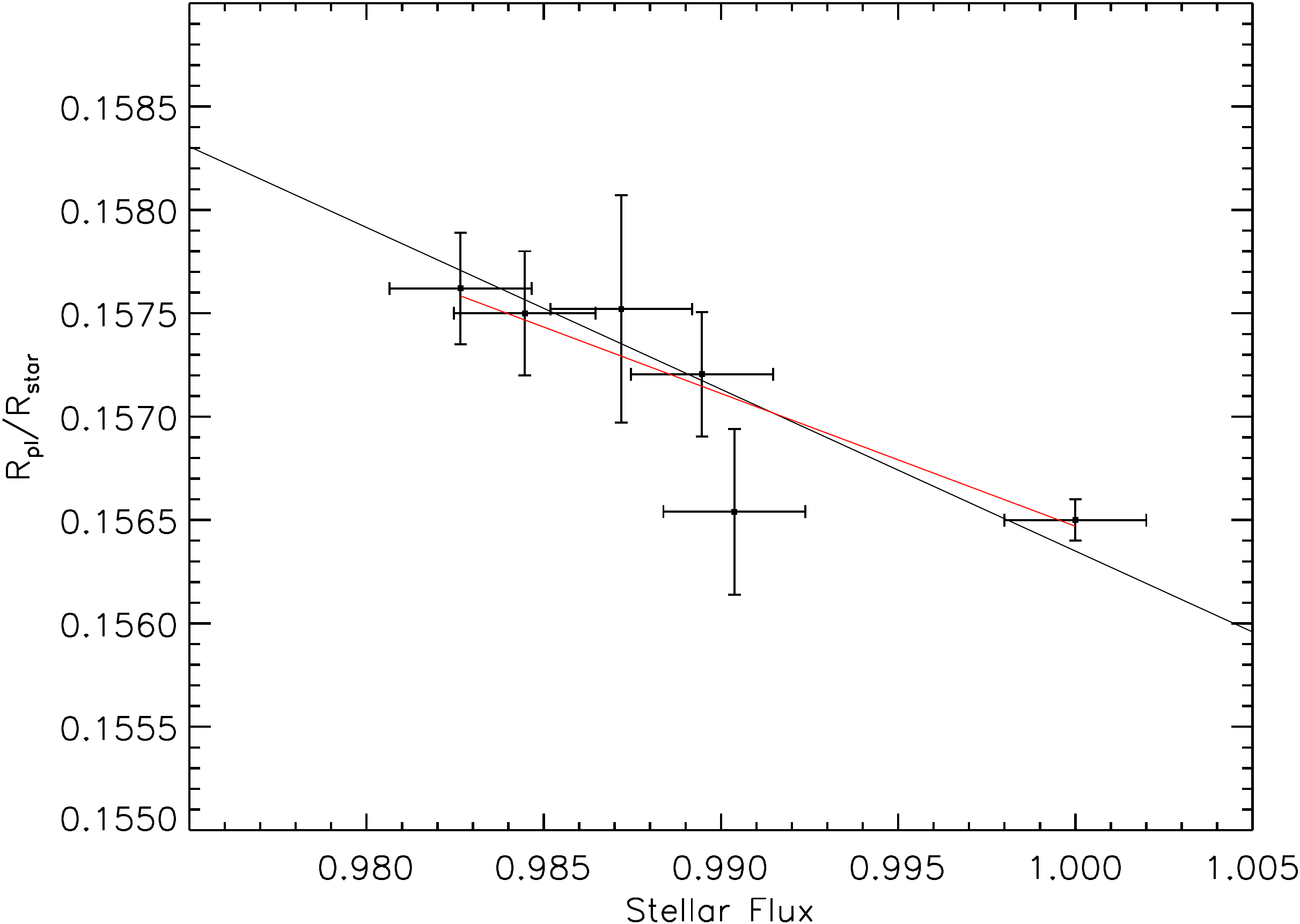}
\caption{HST planet-to-star radius measurements ($R_{pl}/R_{star}$)
   between 5200 and 6100~\AA\ from STIS G430L, STIS G750M and ACS
   HRC plotted against the observed stellar flux at the transit epoch.
   The black solid line is the predicted relation for the radius
   impact from stellar variability, assuming non-occulted dark stellar
   spots.  The red line is the best-fit
 relation from the HST observations, which is in very good agreement with the
 prediction.}
\label{radspotrelation}
\end{figure}
Between the ACS measurements, the two STIS visits presented here, and
three HD189733b transits obtained with the STIS G750M as part of a
separate HST program (GO~11572, PI D. Sing), there are now
sufficiently similar high S/N transit optical observations to
search for the expected correlation in Eq. \ref{EQradiflux} between the measured transit
radius ratio and the activity levels of the star.  To search for this
correlation, we used the 
non-contaminated G430L visit 1 and visit 2 data at the reddest wavelength bin
(5200-5700~\AA), along with the whitelight curves of three similarly high S/N G750M transits
(5880-6200~\AA), estimating the stellar flux level at each transit
epoch from the APT data.  We also used the shortest wavelength ACS radius,
applying a small correction such that the inclination and impact
parameters match those of \cite{2010ApJ...721.1861A}, 
also used for the STIS transits.
As each STIS transit is well sampled temporally, 
we can fit for the radius ratio of each transit using only
occulted-spot free regions. 
The results are plotted in Fig. \ref{radspotrelation}.  A linear fit
between the flux and radius gives as slope of $\Delta
(R_{\rm pl}/R_{\rm star}) / \Delta f(\lambda_0,t)=-$0.064$\pm$0.016, in
excellent agreement with the theoretically predicted slope value of
$-\frac{1}{2}(R_{\rm pl}/R_{\rm star})=-$0.078 from
Eq. \ref{eqradspotrelationd} and \ref{eqradspotrelation}, and the expectation of stellar
variability via dark stellar spots.  
This correlation reinforces the importance and
usefulness of ground-based monitoring for active transiting hosting
stars, and lends credence to the hypothesis that in sufficiently non-active stars,
such stellar activity related radius changes can be negligible
\citep{2011A&A...527A..73S}. 

\begin{figure}
 \includegraphics[width=0.475\textwidth]{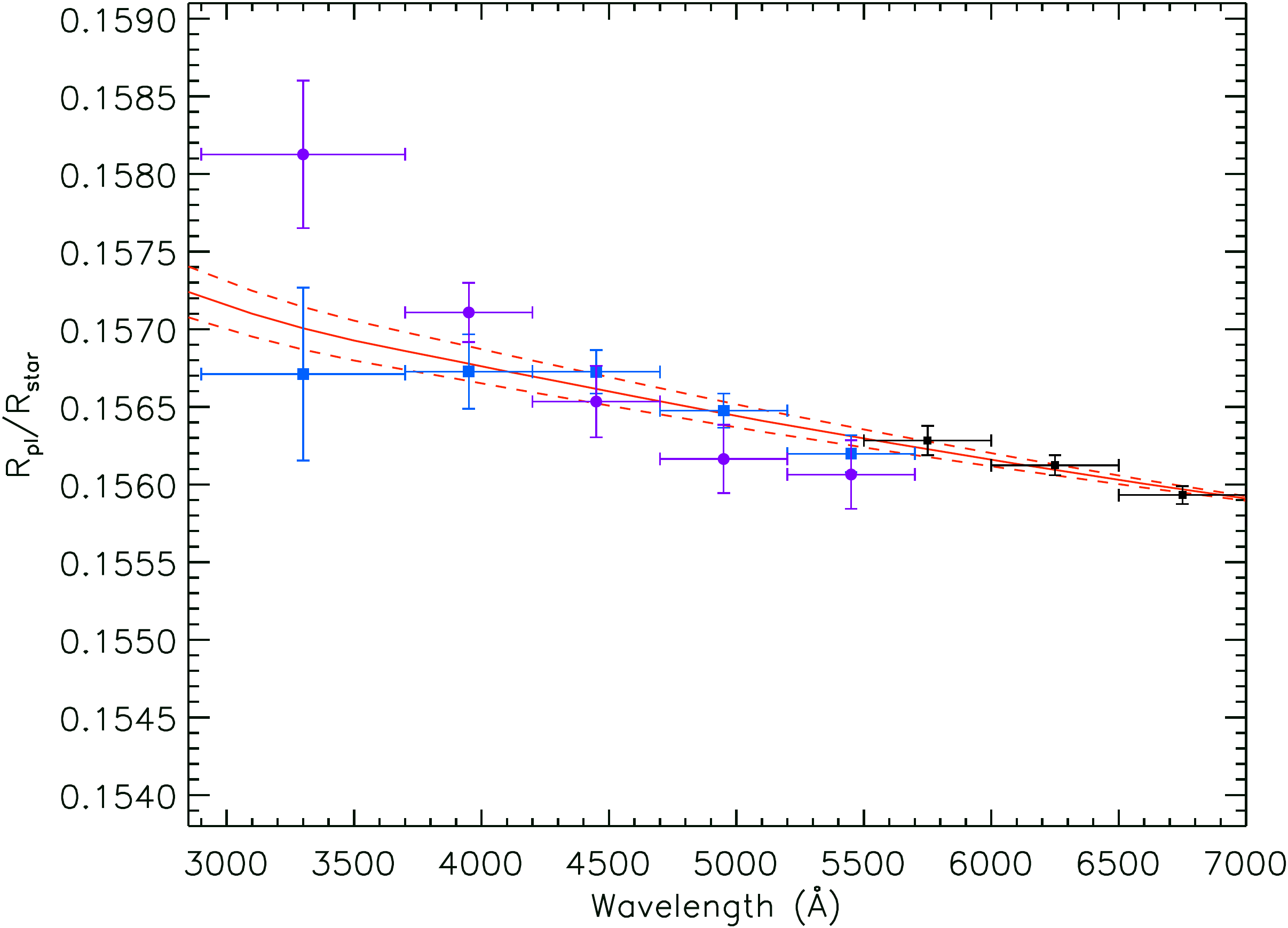}
  \caption{G430L transit radius spectrum for visit 1 (purple circles) and
    visit 2 (blue squares) along with previous ACS measurements (black
    squares) and the prediction from Rayleigh scattering (solid and
    dashed lines).  The wavelength bins are indicated by the X-axis
    error bar size and the 1-$\sigma$ error of the transit fit is indicated by the Y-axis
    error bar size.}
\label{V1V2Rad}
\end{figure}

\begin{figure*}
 \includegraphics[width=14cm]{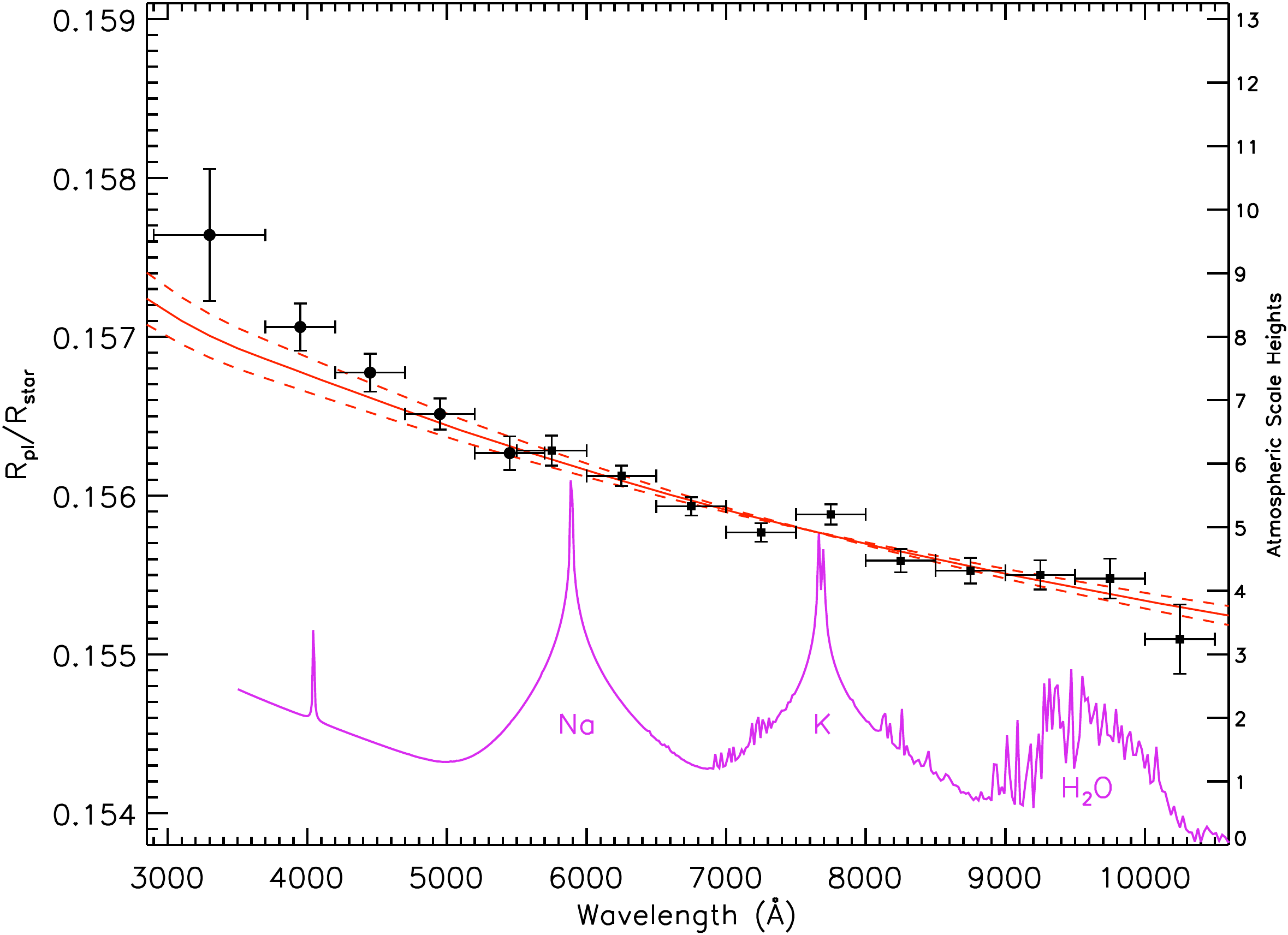}
  \caption{STIS and ACS transmission spectra for HD189733b.  Plotted
    blue-ward of 5600~\AA\ is the STIS G430L measurements (circles)
    with the ACS measurements from Pont et al. (2008) red-ward of
    5600~\AA (squares).  The wavelength bins are indicated by the X-axis
    error bars and the 1-$\sigma$ error is indicated by the Y-axis
    error bars.  The right Y-axis is
    labeled in units of estimated atmospheric scale heights, assuming
    T=1340~K ($H$=0.0004 $R_{pl}/R_{star}$).  The prediction from ACS
    Rayleigh scattering (1340$\pm$150 K red solid and
    dashed lines) is also shown, as is a haze-free model atmosphere
    for HD~189733b from Fortney et al. (2010, Fig. 7) which uses a planet-wide average T-P profile, and is normalized to the radii at infrared wavelengths.}
\label{FigfullSpectra}
\end{figure*}
 
\section{DISCUSSION}
\label{discussion}
We find very good overall agreement between the transmission spectrum of
the two HST visits for HD189733b (see Fig.~\ref{V1V2Rad}, Table
\ref{TableAveRad}).  The largest deviation (at the 2-$\sigma$ level) is
at the very bluest wavelength, which is most sensitive to the
prescriptions used for limb-darkening, as well as the occulted stellar
spot fits of visit 2.  
The average transmission spectrum we obtain (see
Fig.~\ref{FigfullSpectra}) between 4200 and
5700~\AA\ is featureless, lacking the broad Na and K absorption lines, with a blue-ward slope similar to the ACS
measurements.  Like the ACS spectrum (but unlike HD209458b) we find no
evidence for the wide pressure-broadened sodium wings (e.g. \citealt{2010ApJ...709.1396F}), though there is
good evidence that the sodium line core is present both from ground based measurements
\citep{2008ApJ...673L..87R} 
as well as our G750M measurements
(Huitson, Sing, in prep), which indicates that 
either the sodium abundance is much lower in HD189733b or more likely
that the optical and near-UV transmission spectrum covers lower
pressures and higher altitudes than in HD209458b.
The high altitudes are also illustrated by derived planetary
radii for both ACS and STIS, which are both well in excess of those observed in the
near-IR and at Spitzer IRAC wavelengths.   For a hydrostatic 
atmosphere at $\sim$1300~K, the ACS spans
2 scale heights above the 1.88$\mu$m and 8$\mu$m radii of
\cite{2009A&A...505..891S} and \cite{2010ApJ...721.1861A} respectively,
while the G430L spans $\sim$2 to 6 scale heights above.

The featureless slope and lack of the expected sodium and potassium alkali line wings
further indicates optical atmospheric haze, as first detected by
\cite{2008MNRAS.385..109P} using HST ACS. 
The effective transit measured altitude $z$ of a hydrostatic atmosphere as a
function of wavelength $\lambda$ was found in \cite{2008A&A...481L..83L}, 
\begin{equation} 
z(\lambda)=H \mathrm{ln} \left(
    \frac{\varepsilon_{\rm abs}  P_{\rm ref}
      \sigma(\lambda)}{\tau_{\rm eq}}
    \sqrt{\frac{2 \pi R_{\rm pl}}{kT\mu g}} \right), \end{equation}
where 
$\varepsilon_{abs}$ is the abundance of dominating absorbing species, 
$T$ is the atmospheric temperature at $z$,
$H=kT/\mu g$ is the atmospheric scale height,
$\mu$ is the mean mass of the atmospheric particles,
$P_{\rm ref}$ is the pressure at the reference altitude, and
$\sigma(\lambda)$ is the absorption cross section.
This allows the derivation of the apparent planetary radius as a function
of wavelength from the known cross section variations. 
Assuming a scaling law for the cross section in the form
$\sigma=\sigma_0 (\lambda/\lambda_0)^\alpha$, 
the slope of the planet radius as a function
of wavelength is given by
\begin{equation}  
  \frac{dR_{pl}}{d \ln \lambda}= \alpha H =  \alpha \frac{kT}{\mu g}.
\label{slopeeq}
\end{equation}
If the cross section as a function of wavelength is known, the local
atmospheric temperature can thus be estimated by 
\begin{equation}  
\alpha T=\frac{\mu g}{k}\frac{dR_{pl}}{d \ln \lambda}.
\end{equation}
For the ACS measurements,  \cite{2008A&A...481L..83L} 
showed that Rayleigh scattering with $\alpha=-4$ (Rayleigh cross
section is $\sigma=\sigma_0(\lambda/\lambda_0)^{-4}$) provides a
temperature of 1340$\pm$150 K, consistent with other estimates.  They also
concluded that the absorption from atmospheric haze particles is due to
Rayleigh scattering and suggested  MgSiO$_{3}$ as a possible
candidate;  this condensate is indeed predicted to be present in hot Jupiters
and brown dwarfs by \cite{2008A&A...485..547H}. 
MgSiO$_{3}$ is an attractive candidate over many other plausible dust condensates (like Mg$_2$SiO$_4$, MgFeSiO$_4$,
O-deficient silicates, etc.), as the scattering efficiency dominates
over absorption efficiency, giving a Rayleigh scattering profile
when using the Mie approximation.

The STIS G430L measurements between 2900 and 5700~\AA\ gives $\alpha T =$
$-8400\pm2000$~K slope.  
The total magnitude of the
un-occulted spot correction for the visit 1 and visit 2 STIS G430L data is
incorporated in the slope error and accounts for about 40\% of the slope uncertainty.
Assuming that Rayleigh scattering still holds at these shorter wavelengths,
the steeper slope would indicate higher temperatures of 2100$\pm$500~K,
following Eq. \ref{slopeeq}. 
Comparing the two Rayleigh temperatures between the ACS and STIS
datasets, the STIS appears warmer by 760$\pm$522~K which is at the
1.5-$\sigma$ significance level. 

Warmer temperatures at higher
altitudes could indicate the presence of a thermosphere.  Within the
context of Rayleigh scattering, redder wavelengths have a
lower cross section, thus the transmission spectrum would be probing
somewhat deeper
into the atmosphere, toward cooler temperatures below the thermosphere.  A thermosphere at
these high altitudes is expected and is
in agreement with the detection of an escaping exosphere
\citep{2010A&A...514A..72L} 
A similar thermosphere has also been observed at high
altitudes on HD~209458b when using sodium to probe the
temperature profile \citep{2011A&A...527A.110V}.  
Very high temperatures could potentially
pose a problem for Rayleigh scattering via MgSiO$_3$ particles, as they
vaporize near 1400~K,
though the higher temperatures measured in the G430L are only marginally significant.
A Rayleigh scattering dominated atmosphere would imply a high albedo,
as a semi-infinite purely Rayleigh scattering atmosphere has a theoretical
geometric albedo of 0.8 \citep{1974ApJ...192..787P}. 

The larger radii at near-UV wavelengths (blueward of $\sim$4000~\AA) could also be attributed to
other absorbing features, such as a Balmer jump associated with the
escaping atmosphere \citep{2007Natur.445..511B} or sulfur
compounds generated photochemically \citep{2009ApJ...701L..20Z}.
Both scenarios have significant absorption at high altitudes blueward
of $\sim$4000 \AA, though a larger EUV flux for HD~189733b (compared to HD~209458b) should
limit a Balmer jump signature and sulfur compounds have a
characteristically steeper absorption profile.


\section{CONCLUSION}
HD~189733b is now only the second exoplanet to have a full optical
transmission spectrum measured.  Using high S/N observations from the
repaired STIS instrument, we confirm a featureless spectrum with
opacity increasing bluewards in the optical, suggestive of Rayleigh
scattering and atmospheric
haze.  The HST transmission spectral results 
indicate that the atmospheric haze and Rayleigh scattering is an important
feature of HD~189733b's atmosphere
and would imply a high albedo.  
The effect of the haze on the global energy budget of the planet could
be significant, as the wavelength-dependent albedo from Rayleigh
scattering is large, while typical condensate-free hot-Jupiter atmospheres are
dominated by Na and K line wings and have very low optical albedos.
Such a large albedo can be independently checked by optical secondary
eclipse measurements.  For HD~189733b the optical secondary eclipse
has yet to be measured, as the stellar activity makes it difficult to build up a detection
from multiple eclipse events \citep{2006ApJ...646.1241R}. 

The list of key differences between the two well studied
hot Jupiters is growing, with HD~209458b likely featuring a
low albedo, stratospheric temperature inversion, inflated planetary
radius, and large Na alkali line wings while HD~189733b has a
high albedo, no stratospheric temperature inversion, a non-inflated
planetary radius, and a global haze covering significant alkali line
wing absorption.  Such differences points toward a large diversity 
between the broader class of hot-Jupiter atmospheres, as HD~209458b and
HD~189733b only differ by a few hundred degrees K.

\section*{Acknowledgments}
We would like to thank the astronauts for their courageous work
during the HST Servicing Mission 4.
We also thank Jonathan Fortney for providing his model atmospheres,
Gilda Ballester for helpful detailed comments, and our reviewer Ignas Snellen for constructive commentary.
This work is based on observations
with the NASA/ESA Hubble Space Telescope, obtained at the
Space Telescope Science Institute (STScI) operated by AURA, Inc.  
Support for this work was provided by NASA through the GO-11740.01-A
grant from the STScI. We wish to acknowledge the support of a STFC Advanced Fellowship (FP).
W.H. acknowledges support by the European Research Council under the
European Community's 7th Framework Programme (FP7/2007-2013 Grant
Agreement no. 247060).

\bibliographystyle{mn2e} 
\bibliography{HD189.STIS} 

\end{document}